\documentclass[aps,prd,nofootinbib,superscriptaddress,noshowkeys,10pt]{revtex4-2}
\usepackage[utf8]{inputenc}
\usepackage[english]{babel}
\usepackage{graphicx,wrapfig,subfigure,float}
\usepackage{amsfonts,amsmath,amssymb,amstext}
\usepackage{color}

\newcommand{\slsh}[1]{\not \! #1}

\begin{document}

\title{Scalar boson emission from a magnetized relativistic plasma}

\author{Jorge Jaber-Urquiza}
\email{jorgejaber@ciencias.unam.mx}
\affiliation{Facultad de Ciencias, Universidad Nacional Aut\' onoma de M\' exico, Apartado Postal 50-542, Ciudad de M\'exico 04510, M\'exico}
\affiliation{College of Integrative Sciences and Arts, Arizona State University, Mesa, Arizona 85212, USA}

\author{Igor A. Shovkovy}
\email{igor.shovkovy@asu.edu}
\affiliation{College of Integrative Sciences and Arts, Arizona State University, Mesa, Arizona 85212, USA}
\affiliation{Department of Physics, Arizona State University, Tempe, Arizona 85287, USA}

\date{October 31, 2023}


\begin{abstract}
We investigate the differential emission rate of neutral scalar bosons from a highly magnetized relativistic plasma. We show that three processes contribute at the leading order: particle splitting ($\psi\rightarrow \psi+\phi $), antiparticle splitting ($\bar{\psi} \rightarrow \bar{\psi}+\phi $), and particle-antiparticle annihilation ($\psi + \bar{\psi}\rightarrow \phi $). This is in contrast to the scenario with zero magnetic field, where only the annihilation processes contribute to boson production. We examine the impact of Landau-level quantization on the energy dependence of the rate and investigate the angular distribution of emitted scalar bosons. The differential rate resulting from both (anti)particle splitting and annihilation processes are typically suppressed in the direction of the magnetic field and enhanced in perpendicular directions. Overall, the background magnetic field significantly amplifies the total emission rate. We speculate that our model calculations provide valuable theoretical insights with potentially important applications.
\end{abstract}

\maketitle

\section{Introduction}
\label{sec.intro}

The properties of matter under extreme conditions, where relativistic effects play a profound role, are a source of great fascination. This fascination is not surprising, as such extreme conditions naturally occur in the early Universe and in stars~\cite{Raffelt:1996wa,Borisov:1997zm,Harding:2006qn,Potekhin:2010aii,Granot:2015xba}. However, replicating these conditions in a laboratory setting is exceptionally challenging. The most promising efforts in this direction involve heavy-ion experiments conducted at the Relativistic Heavy Ion Collider (RHIC) at Brookhaven and the Large Hadron Collider (LHC) at CERN \cite{Heinz:2013th,Bzdak:2019pkr}. In these experiments, heavy ions are accelerated to sufficiently high energies to produce tiny volumes of quark-gluon plasma (QGP) \cite{Stoecker:1986ci,Gyulassy:2004zy}. Although this new state of matter has a very brief lifetime and is likely far from equilibrium, some of its properties can still be deduced \cite{STAR:2005gfr,PHENIX:2004vcz,PHOBOS:2004zne}.

Over the last two decades, significant progress has been made in understanding the properties of hot QGP \cite{Braun-Munzinger:2015hba,Busza:2018rrf}. The emerging picture can be summarized as follows. Heavy-ion collisions generate matter with high energy density, which is initially far from equilibrium. Due to strong interaction, this matter rapidly approaches a quasi-equilibrium QGP state. Furthermore, it behaves almost like a perfect hydrodynamic fluid, undergoing expansion, cooling, and eventual hadronization ~\cite{Ryu:2015vwa,Denicol:2015nhu,Tang:2011xq,Petran:2013lja}. The resulting hadrons carry its remnants to the detectors, enabling one to unveil the  properties of hot QGP.

The QGP produced in heavy-ion collisions not only possesses an extremely high temperature but also carries a strong magnetic field \cite{Skokov:2009qp,Voronyuk:2011jd,Deng:2012pc,Bloczynski:2012en} and exhibits high vorticity \cite{STAR:2017ckg,STAR:2018gyt,STAR:2019erd}. Theoretical investigations indicate that both the magnetic field and vorticity can modify the observed properties of QGP~\cite{Liao:2014ava,Miransky:2015ava,Kharzeev:2015znc,Huang:2020xyr,Shovkovy:2021yyw}. Of particular significance are the observables linked to electromagnetic probes, as they convey information about the plasma's properties across all stages of its evolution~\cite{Qiu:2011iv}. 

In this paper, we will study the differential production rate of neutral scalar bosons within a strongly magnetized relativistic plasma. Previously, an attempt to address this problem was undertaken in Ref.~\cite{Bandyopadhyay:2018gbw}. However, only simplified kinematics with momenta of scalar bosons parallel to the magnetic field was considered. Another related study on the scalar boson decay at zero temperature and weak field was reported in Ref.~\cite{Jaber-Urquiza:2018oex}. In both instances, the constraints imposed by kinematics allowed only for the contribution of particle-antiparticle annihilation processes to the absorptive part of the self-energy (or boson decay). Herein, we undertake a comprehensive approach, removing all constraints on kinematics, permitting arbitrary magnetic field strengths, and incorporating the thermal effects of the plasma to address this problem in its entirety.

At first glance, this problem may not have direct phenomenological implications for QGP in heavy-ion collisions. After all, there are no known spin-zero particles to be emitted from a relativistic plasma. Nevertheless, we believe that this problem has theoretical value. By comparing the results with the emission of photons~\cite{Yee:2013qma,Tuchin:2014pka,Zakharov:2016mmc,Wang:2020dsr,Wang:2021ebh} and dileptons \cite{Sadooghi:2016jyf,Bandyopadhyay:2016fyd,Bandyopadhyay:2017raf,Ghosh:2018xhh,Islam:2018sog,Das:2019nzv,Ghosh:2020xwp,Chaudhuri:2021skc,Das:2021fma,Wang:2022jxx} (i.e., spin-one channel) studied previously, we can get insights into the impact of particle spin on emission rates and angular distributions. This hypothetical scenario also extends our understanding of the fundamental laws of physics and their potential applications in other fields. 

For example, neutral scalar bosons often appear in dark matter \cite{Burgess:2000yq,Buckley:2014fba,Hui:2016ltb} and inflationary models \cite{Mukhanov:2005sc}. Moreover, their properties, when modified by a nonzero temperature and primordial magnetic fields, can have cosmological implications \cite{Piccinelli:2014dya,Piccinelli:2021vbq}. In the end, our goal is to refine our theoretical tools and expand the frontiers of scientific knowledge. Not every thought experiment or hypothetical scenario leads to a discovery, but more often than not, it provides fresh insights and perspectives.

The paper is organized as follows. We introduce the model of magnetized plasma with a single flavor of fermion species, coupled to a neutral scalar field via a Yukawa-type interaction, in Sec.~\ref{sec.model}. There, we also define the differential emission rate in terms of the imaginary part of the scalar-boson self-energy. The general expression for the self-energy at nonzero temperature is obtained in Sec.~\ref{sec.calculations}. In the derivation, we utilize the Landau-level representation for the fermion propagator, which allows us to extract an analytical expression for the imaginary part of the self-energy in the form of a convergent series.  In Sec.~\ref{sec.num}, the corresponding result is used to calculate the differential emission rate of scalar bosons from a magnetized plasma. We study in detail the angular dependence of the emission rate, as well as analyze the partial contributions due to annihilation (i.e., $\psi+\bar{\psi}\to \phi$) and splitting (i.e., $\bar{\psi}\to \bar{\psi}+\phi$ and $\psi\to \psi+\phi$) processes. A discussion of the main findings and a summary of the results are given in Sec.~\ref{sec.concl}. For comparison, the bosonic self-energy in the zero magnetic field limit is presented in Appendix~\ref{app.IntegrationB=0}. 

\section{Model}
\label{sec.model}

 For simplicity, we consider a model of magnetized plasma with a single flavor of fermion species $\psi$. By assumption, the fermions interact with the neutral scalar field $\phi$ via a Yukawa interaction. The corresponding Lagrangian density reads
\begin{equation}
 \mathcal{L}=\bar{\psi}\left(i\gamma^{\mu} D_{\mu}-m\right)\psi+\frac{1}{2}\partial^\mu\phi\partial_\mu\phi-\frac{1}{2} M^2\phi^2-g\phi\bar{\psi}\psi,
 \label{eq.lagrangian}
\end{equation}
where  $m$ and $M$ are the masses of the fermion and scalar particles, and $q$ is the electric charge of the fermion. The covariant derivative is defined as usual, i.e., $D^\mu\equiv\partial^\mu+iqA^{\mu}(x)$, where $A^{\mu}(x)$ is an Abelian gauge field, capturing the effect of a background magnetic field $\mathbf{B}$. The corresponding field strength tensor is given by $F^{\mu\nu}=\partial^\mu A^{\nu}(x)-\partial^\nu A^{\mu}(x)$. Without loss of generality, we will assume that the magnetic field points along the $z$ axis and use the following Landau gauge: $A^{\mu}(x) =- y B\delta^{\mu}_{1}$. The explicit form of the strength tensor reads $F^{\mu\nu}=-\varepsilon^{0\mu\nu3}B$. Here we use the conventional definition of the contravariant coordinates, i.e., $x^\mu=(t,x,y,z)$, and the Minkowski metric $g_{\mu\nu} =\mbox{diag}(1,-1,-1,-1)$.

\begin{figure}[t]
\includegraphics[width=0.25\textwidth]{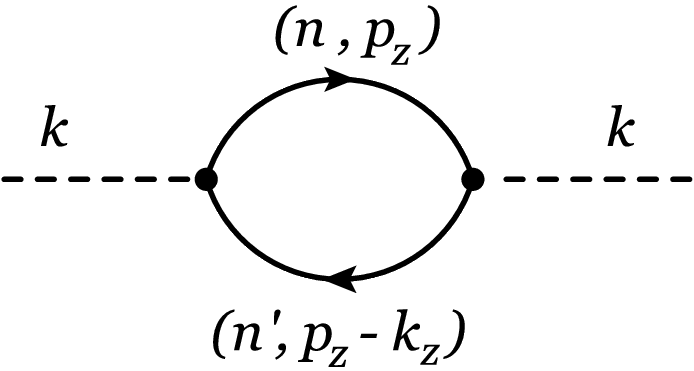}
\caption{Leading order one-loop Feynman diagram for the scalar self-energy.}
\label{fig.selfenergy}
\end{figure}

The differential thermal emission rate of scalar bosons from the corresponding plasma is given by
\begin{equation}
 \frac{d^3 R}{d^3 k}=-\frac{n_B(\Omega)}{(2\pi)^3\Omega} \mbox{Im}\left[\Sigma^R(\Omega,\mathbf{k})\right],
 \label{eq:diff-rate}
\end{equation}
where $\Omega=\sqrt{M^2+|\mathbf{k}|^2}$ is the (on shell) scalar particle energy,  $n_B(\Omega)=1/\left[e^{\Omega/T}-1\right]$ is the Bose-Einstein distribution function, and $\Sigma^R(\Omega,\mathbf{k})$ is the retarded self-energy of the scalar field. At leading order in coupling, the latter is determined by the one-loop Feynman diagram in Fig.~\ref{fig.selfenergy}, where the solid and dashed lines represent fermions and bosons, respectively. Because of the background magnetic field, the fermion propagators are labeled by the longitudinal momenta and the Landau-level indices.

Note that, in view of the detailed balance, the expression in Eq.~(\ref{eq:diff-rate}) can represent either the emission or the absorption rate per unit volume. However, the total emission (or absorption) rate can be also affected by the system size, if the latter is comparable to or larger than the mean free path $l_{\phi}$ of the scalar bosons with energy $\Omega$. For simplicity, we will ignore the corresponding effects below. If needed, however, they could be incorporated approximately by separating the surface layers of depth $l_{\phi}$ from the rest of the plasma. The rate in Eq.~(\ref{eq:diff-rate}) is valid only for the surface layers. The emission from the inner parts is approximately vanishing.  

In view of the rotational symmetry of a magnetized plasma about the magnetic field direction, the differential rate is independent of the azimuthal angle $\phi$ (which is measured in $xy$-plane from the positive $x$-axis). Taking this fact into account, we derive the following expression for the total rate integrated over all directions:
\begin{equation}
 \frac{dR}{dk}=-\int_{0}^{\pi} \frac{k^2 n_B(\Omega)}{(2\pi)^2\Omega} \mbox{Im}\left[\Sigma^R(\Omega,\mathbf{k})\right] \sin\theta d\theta ,
 \label{eq:diff-rate.dk}
\end{equation}
where the polar angle $\theta$ is measured from the positive $z$-axis towards the $xy$-plane. In other words, the transverse and the longitudinal components of the boson momentum are $k_\perp = k\sin{\theta}$ and $k_z = k\cos{\theta}$, respectively. 
By rewriting the rate in terms of the boson energy, we have 
\begin{equation}
 \frac{dR}{d\Omega}=-\int_{0}^{\pi} \frac{k\, n_B(\Omega)}{(2\pi)^2} \mbox{Im}\left[\Sigma^R(\Omega,\mathbf{k})\right] \sin\theta d\theta .
 \label{eq:diff-rate.dOmega}
\end{equation}
In order to characterize the angular profile of emission, we will utilize the following definition of the ellipticity parameter:
\begin{equation}
v_2 = - \frac{\int_{0}^{\pi} \left(d^3 R/ d^3 k\right) \cos(2\theta) d\theta}{\int_{0}^{\pi} \left(d^3 R/ d^3 k\right) d\theta},
\label{eq:v2}
\end{equation}
which is analogous to the one used in heavy-ion physics but expressed in terms of a different angular coordinate. An extra negative sign in the definition ensures that a positive value of ellipticity ($v_2 >0$) describes an oblate emission profile, i.e., stronger average emission in the directions perpendicular to the magnetic field (or, in heavy-ion physics language, in the reaction plane). A negative value of ellipticity ($v_2 <0$) implies a prolate emission profile, i.e., stronger average emission in the directions parallel to the magnetic field (or, in heavy-ion physics language, perpendicularly to the reaction plane).
 
\section{One-loop self-energy}
\label{sec.calculations}

 In the presence of a background magnetic field, translation symmetry is broken in the plane perpendicular to the magnetic field. As a consequence, the transverse momenta are not good quantum numbers for classifying fermionic states. This fact is also reflected in the structure of the fermion propagator, which takes the following form in coordinate space \cite{Schwinger:1951nm}:
\begin{equation}
 S(x,y)=\exp\left(-iq\int_{y}^{x} A_\mu(x)dx^\mu\right) \bar{S}(x-y),
 \label{eq.qB-PropConf}
\end{equation}
where the first factor is the so-called Schwinger's phase. Formally, this phase is the only part that breaks the translation symmetry. The second factor, $\bar{S}(x-y)$, is a translation invariant part of the propagator. Its explicit form will be given below.

In coordinate space, the one-loop self-energy of the scalar field is given by 
\begin{equation}
\Sigma (x-y)=i g^2 \mbox{Tr}\left[\bar{S}(x-y) \bar{S}(y-x)\right],
\label{eq.self-enegry}
\end{equation}
see Fig.~~\ref{fig.selfenergy}, where the trace runs over the Dirac indices. Note that, at this leading order in coupling, it is determined only by the translation invariant part of the fermion propagator $\bar{S}(x-y)$. 

It should not be surprising that the dependence of $\bar{S}(x)$ on the transverse and longitudinal spatial coordinates (i.e., $\mathbf{r}_\perp$ and  $z$, respectively) is very different. Unlike translations in the $xy$-plane, translations in the $z$ direction are part of the remaining symmetry in the problem. In other words, the corresponding longitudinal momentum $k_z$ is a good quantum number. Thus, it convenient to use the following mixed representation for translation invariant part of the propagator:
\begin{equation}
 \bar{S}(x)= \int \frac{d^2 p_\parallel }{(2\pi)^2} \tilde{S}(p_\parallel ; \mathbf{r}_\perp)e^{-ip_\parallel\cdot x_\parallel},
 \label{eq.prop-p-r}
\end{equation}
where, by definition, $x_\parallel = (t,z)$, $\mathbf{r}_\perp = (x,y)$, and $p_\parallel =(p_0,p_z)$ is the longitudinal momentum. The explicit form of $\tilde{S}(p_\parallel ; \mathbf{r}_\perp)$ reads \cite{Miransky:2015ava}
\begin{equation}
 \tilde{S}(p_\parallel ; \mathbf{r}_\perp)= i \frac{e^{-\zeta /2}}{2\pi \ell^2} \sum_{n=0}^{\infty}\frac{\tilde{D}\left(p_\parallel;\mathbf{r}_\perp\right)}{p_\parallel^2-m^2-2n|qB|}
 \label{eq.MF-PropaMixed1}
\end{equation}
with the shorthand notation $\zeta \equiv|\mathbf{r}_\perp|^2/(2\ell^2)$ and
\begin{equation}
 \tilde{D}\left(p_\parallel;\mathbf{r}_\perp\right)\equiv\left(\slsh{p}_\parallel+m\right)
 \left[\mathcal{P}_+L_n\left(\zeta \right)+\mathcal{P}_-L_{n-1}\left(\zeta \right)\right]-i\frac{\slsh{\mathbf{r}}_\perp}{\ell^2}L^1_{n-1}\left(\zeta \right),
 \label{eq.MF-PropaMixed2}
\end{equation}
where $\ell\equiv1/\sqrt{|qB|}$ is the magnetic length, $L_n(z)$ are the Laguerre polynomials,
$L^a_n(z)$ are the generalized Laguerre polynomials, and $\mathcal{P}_{\pm}\equiv\frac{1}{2}\left(1\pm i\text{sign}\left(qB\right)\gamma^1\gamma^2\right)$ are the spin projectors along the magnetic field direction.  

After substituting the expression for $\bar{S}(x)$ into the definition of self-energy in Eq.~(\ref{eq.self-enegry}) and performing the Fourier transform, we derive the following momentum representation:
\begin{equation}
\Sigma (k)=i g^2 \int \frac{d^2 p_\parallel}{(2\pi)^2} \int d^2 \mathbf{r}_\perp e^{-i\mathbf{r}_\perp\cdot\mathbf{k}_\perp}
\mbox{Tr}\left[\tilde{S}(p_\parallel; \mathbf{r}_\perp) \tilde{S}(p_\parallel-k_\parallel;-\mathbf{r}_\perp)\right].
\label{eq.self-enegry-momentum}
\end{equation}
By using the fermion propagator in Eq.~(\ref{eq.MF-PropaMixed1}) and performing the trace over the Dirac indices, we obtain the following expression for the scalar self-energy:
\begin{eqnarray}
 \Sigma (k)&=&-\frac{ig^2}{2\pi^2\ell^4} \int\frac{d^{2}p_\parallel}{(2\pi)^{2}}\int d^2\mathbf{r}_\perp 
 e^{-i\mathbf{r}_\perp\cdot\mathbf{k}_\perp} e^{-\zeta } \nonumber \\
&\times& \sum_{n,n^{\prime}}\frac{\left(m^2+p_\parallel\cdot\left(p-k\right)_\parallel\right)\left[L_{n}(\zeta )L_{n^{\prime}}(\zeta )+L_{n-1}(\zeta )L_{n^{\prime}-1}(\zeta )\right]-\frac{2\mathbf{r}_\perp^2}{\ell^4} L_{n-1}^1(\zeta )L_{n^{\prime}-1}^1(\zeta )}{\left(p_\parallel^2-m^2-2n|qB|\right)\left((p_\parallel-k_\parallel)^2-m^2-2n^{\prime}|qB|\right)}.
 \label{eq.MF-VEdiagA6}
\end{eqnarray}
The integration over the transverse spatial coordinates can be performed exactly using the same approach as in Refs.~\cite{Wang:2020dsr,Wang:2021ebh}. The result reads 
\begin{equation}
  \Sigma (k)=-i\frac{g^2}{\pi\ell^2}\int\frac{d^{2}p_\parallel}{(2\pi)^2}\sum_{n,n^{\prime}}\frac{\left(m^2+p_\parallel\cdot\left(p-k\right)_\parallel\right)\left(\mathcal{I}^{n,n^{\prime}}_0(\xi)+\mathcal{I}^{n-1,n^{\prime}-1}_0(\xi)\right)-\frac{2}{\ell^2}\mathcal{I}^{n-1,n^{\prime}-1}_2(\xi)}{\left(p_\parallel^2-m^2-2n|qB|\right)\left((p_\parallel-k_\parallel)^2-m^2-2n^{\prime}|qB|\right)}.
 \label{eq.MF-VEdiagA8}
\end{equation}
where $\xi \equiv (k_\perp\ell)^2/2$ and the two new functions are
\begin{eqnarray}
  \label{eq.I0int}
 \mathcal{I}_0^{n,n^{\prime}}(\xi)&\equiv &   (-1)^{n+n^{\prime}}e^{-\xi}L_n^{n^{\prime}-n}(\xi)L_{n^{\prime}}^{n-n^{\prime}}(\xi),\\
 \mathcal{I}_2^{n,n^{\prime}}(\xi)&\equiv & 2(n^{\prime}+1)(-1)^{n+n^{\prime}}e^{-\xi}L_n^{n^{\prime}-n}(\xi)L_{n^{\prime}+1}^{n-n^{\prime}}(\xi).
  \label{eq.I2int}
\end{eqnarray}
To take thermal effects into account, we introduce the Matsubara frequencies through the imaginary time formalism. Then, replacing the fermion energy $p_0\to i\omega_k=2i \pi(k+1)T$ and the boson energy with the bosonic counterpart, i.e., $k_0\to i\Omega_m=2i \pi m T$, the corresponding finite-temperature scalar self-energy reads
\begin{equation}
  \Sigma (i\Omega_m,\mathbf{k})=\frac{g^2T}{\pi\ell^2} \sum_{k=-\infty}^{\infty}\int\frac{dp_z}{2\pi}\sum_{n,n^{\prime}}\frac{\left(m^2+p_\parallel\cdot\left(p-k\right)_\parallel\right)\left(\mathcal{I}^{n,n^{\prime}}_0(\xi)+\mathcal{I}^{n-1,n^{\prime}-1}_0(\xi)
   \right)-\frac{2}{\ell^2}\mathcal{I}^{n-1,n^{\prime}-1}_2(\xi)}{\left((i\omega_k)^2-p_z^2-m^2-2n|qB|\right)\left((i\omega_k-i\Omega_m)^2-(p_z-k_z)^2-m^2-2n^{\prime}|qB|\right)},
 \label{eq.MF-VEdiagA9}
\end{equation}
where the shorthand notation $p_\parallel\cdot\left(p-k\right)_\parallel$ stands for $ i\omega_k(i\omega_k-i\Omega_m)-p_z(p_z-k_z)$. Computing the sum over the Matsubara frequencies, we derive the following expression for the self-energy:
\begin{eqnarray} 
\Sigma (i\Omega_m,\mathbf{k}) &=& \frac{g^2}{\pi\ell^2}\int \frac{dp_z}{2\pi}\sum_{n,n^{\prime}}\sum_{\eta,\lambda=\pm1}
\frac{n_F\left(E_{n,p_z}\right)-n_F\left(\lambda E_{n^{\prime},p_z-k_z}\right)}{4\lambda E_{n,p_z}E_{n^{\prime},p_z-k_z}\left(E_{n,p_z}-\lambda E_{n^{\prime},p_z-k_z}+i\eta\Omega_m\right)}\nonumber \\
  &\times&\bigg[\left(\lambda E_{n,p_z}E_{n^{\prime},p_z-k_z}+m^2-p_z\left(p_z-k_z\right)\right)\left(\mathcal{I}^{n,n^{\prime}}_0(\xi)+\mathcal{I}^{n-1,n^{\prime}-1}_0(\xi)\right)-\frac{2}{\ell^2}\mathcal{I}^{n-1,n^{\prime}-1}_2(\xi)\bigg],
 \label{eq.MF-VEdiagA10}
\end{eqnarray}
where $E_{n,p_z}\equiv\sqrt{p_z^2+m^2+2n|qB|}$ and $E_{n^{\prime},p_z-k_z}\equiv\sqrt{(p_z-k_z)^2+m^2+2n^{\prime}|qB|}$ are the Landau level energies, and $n_F(\Omega)=1/\left(e^{\Omega/T}+1\right)$ is the Fermi-Dirac distribution function.
In the derivation we used the following general result: 
\begin{equation}
 T\sum_{k=-\infty}^{\infty}\frac{i\omega_k(i\omega_k-i\Omega_m)\mathcal{Y}+\mathcal{Z}}{\left[(i\omega_k)^2-a^2\right]\left[(i\omega_k-i\Omega_m)^2-b^2\right]}=\sum_{\eta,\lambda=\pm1}\frac{n_F(a)-n_F(\lambda b)}{4\lambda ab\left(a-\lambda b+\eta i\Omega_m\right)}\left[\lambda a b\mathcal{Y}+\mathcal{Z}\right].
 \label{eq.temp.matsubarasum}
\end{equation}

To obtain the self-energy in Minkoswky space, we need to perform a suitable analytic continuation in Eq.~(\ref{eq.MF-VEdiagA10}). The retarded expression for the self-energy is obtained by replacing $i\Omega_m\to\Omega+i\epsilon$
\begin{eqnarray}
  \Sigma^R (\Omega,\mathbf{k}) &=& \frac{g^2}{\pi\ell^2}\int \frac{dp_z}{2\pi}\sum_{n,n^{\prime}}\sum_{\eta,\lambda=\pm1}\frac{n_F\left(E_{n,p_z}\right)-n_F\left(\lambda E_{n^{\prime},p_z-k_z}\right)}{4\lambda E_{n,p_z}E_{n^{\prime},p_z-k_z}\left(E_{n,p_z}-\lambda E_{n^{\prime},p_z-k_z}+\eta\Omega+i\eta\epsilon\right)} \nonumber \\
  &\times&\bigg[\left(\lambda E_{n,p_z}E_{n^{\prime},p_z-k_z}+m^2-p_z\left(p_z-k_z\right)\right)\left(\mathcal{I}^{n,n^{\prime}}_0(\xi)+\mathcal{I}^{n-1,n^{\prime}-1}_0(\xi)\right)-\frac{2}{\ell^2}\mathcal{I}^{n-1,n^{\prime}-1}_2(\xi)\bigg],
 \label{eq.MF-VEdiagA11}
\end{eqnarray}
where $\epsilon \to +0$. 

It should be noted that the expression for the self-energy in Eq.~(\ref{eq.MF-VEdiagA11}) contains both vacuum and thermal contributions. While the latter is finite, the former has an ultraviolet divergence. Therefore, one has to regularize it in order to proceed with the calculation. Fortunately, only the real part of the self-energy is divergent. The imaginary part, which appears in the definition of the emission rate, is finite. 

\subsection{Absorptive part of the self-energy}
\label{subsec:self-energy}

 From the expression for the retarded self-energy in Eq.~(\ref{eq.MF-VEdiagA11}), one can extract the imaginary part by using the well-known  Sokhotski formula, see Eq.~(\ref{eq:Sokhotski}). The corresponding result reads
\begin{eqnarray}
  \mbox{Im}\left[\Sigma^R (\Omega,\mathbf{k})\right] &=& -\frac{g^2}{\ell^2}\int \frac{dp_z}{2\pi}\sum_{n,n^{\prime}}\sum_{\eta,\lambda=\pm1}\frac{n_F\left(E_{n,p_z}\right)-n_F\left(\lambda E_{n^{\prime},p_z-k_z}\right)}{4\eta\lambda E_{n,p_z}E_{n^{\prime},p_z-k_z}}\delta\left(E_{n,p_z}-\lambda E_{n^{\prime},p_z-k_z}+\eta\Omega\right)\nonumber\\
  &\times& \bigg[\left(\lambda E_{n,p_z}E_{n^{\prime},p_z-k_z}+m^2-p_z\left(p_z-k_z\right)\right)\left(\mathcal{I}^{n,n^{\prime}}_0(\xi)+\mathcal{I}^{n-1,n^{\prime}-1}_0(\xi)\right)-\frac{2}{\ell^2}\mathcal{I}^{n-1,n^{\prime}-1}_2(\xi)\bigg].
 \label{eq.MF-VEdiagA12}
\end{eqnarray}
Note that the Dirac $\delta$-function inside the integrand enforces the following energy conservation relation:
\begin{equation}
 E_{n,p_z}-\lambda E_{n^{\prime},p_z-k_z}+\eta\Omega=0.
 \label{eq.EnergyConservation}
\end{equation}
The imaginary part of the self-energy (\ref{eq.MF-VEdiagA12}) is an odd function of the scalar energy $\Omega$. Without loss of generality, therefore, we will assume that $\Omega>0$ from this point onward.

\begin{figure}[t]
  \subfigure[]{\includegraphics[width=0.22\textwidth]{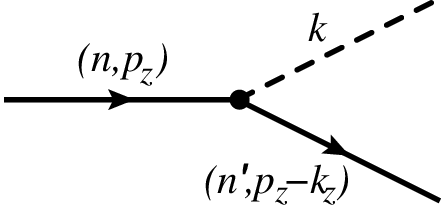}}
  \hspace{0.12\textwidth}
  \subfigure[]{\includegraphics[width=0.22\textwidth]{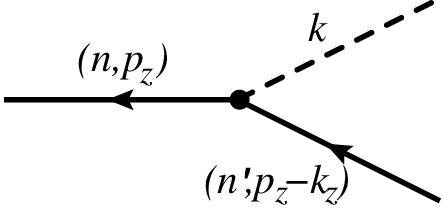}}
  \hspace{0.12\textwidth}
  \subfigure[]{\includegraphics[width=0.22\textwidth]{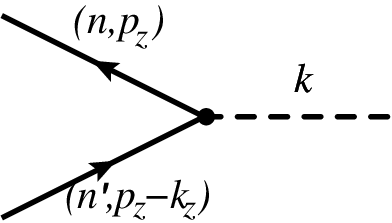}}
\caption{Feynman diagrams for the three processes involving a scalar boson and fermion states in the Landau levels $n$ and $n^{\prime}$: (a) particle splitting $\psi\to \psi+\phi$, (b) antiparticle splitting $\bar{\psi}\to\bar{\psi}+\phi$,
(c) particle-antiparticle annihilation $\psi+\bar{\psi}\to\phi$.}
\label{fig.FeyDiagProcesses}
\end{figure}

Depending on the choice of signs of $\lambda$ and $\eta$, the energy conservation relation (\ref{eq.EnergyConservation}) represents one of the three possible processes involving particles and/or antiparticles states with Landau indices $n$ and $n^{\prime}$. Two of them correspond to particle-splitting processes involving fermions ($\lambda=1$ and $\eta=-1$) or antifermions ($\lambda=1$ and $\eta=-1$). In Fig.~\ref{fig.FeyDiagProcesses}, they are represented by the diagrams in panels (a) and (b), respectively. The third process ($\lambda=-1$ and $\eta=-1$) corresponds to the fermion-antifermions annihilation, represented by the diagram in panel (c) of Fig.~\ref{fig.FeyDiagProcesses}. When $\Omega$ is positive, there are no physical processes associated with the fourth combination of signs, i.e., $\lambda=-1$ and $\eta=1$. It is clear since the energy conservation equation (\ref{eq.EnergyConservation}) has no real solutions in this case.

The necessary and sufficient conditions for having real-valued solutions to the energy conservation equation (\ref{eq.EnergyConservation}) are given as follows:
\begin{eqnarray}
\psi\to\psi+\phi:&\qquad&\sqrt{\Omega^2-k_z^2}\leq k_{-} \ \mbox \ \text{and} \ \mbox \ n>n^{\prime},
\label{process1}
\\
\bar{\psi}\to\bar{\psi}+\phi:&\qquad&\sqrt{\Omega^2-k_z^2}\leq k_{-} \ \mbox \ \text{and} \ \mbox \ n<n^{\prime},
\label{process2}
\\ 
\psi+\bar{\psi}\to\phi:&\qquad&\sqrt{\Omega^2-k_z^2}\geq k_{+} ,
\label{process3}
\end{eqnarray}
for the three types of processes. Here we introduced the following shorthand notation for the transverse momenta thresholds: 
\begin{equation}
 \label{eq.kminusplus}
 k_{\pm}\equiv\bigg|\sqrt{m^2+2n|qB|}\pm\sqrt{m^2+2n^{\prime}|qB|}\bigg|,
\end{equation}
which depend on the Landau-level indices $n$ and $n^{\prime}$. The constraints for $\Omega$ are identical for the two particle-splitting processes in Eqs.~(\ref{process1}) and (\ref{process2}), except for the restrictions on the Landau-level indices. However, they are very different from the kinematic constraint for the annihilation process in Eq.~(\ref{process3}). The requirements $n>n^{\prime}$ and $n<n^{\prime}$ in Eqs.~(\ref{process1}) and (\ref{process2}), respectively, ensure that the initial Landau level state lies above the final one. 

By solving the energy conservation relation (\ref{eq.EnergyConservation}), we find the following pair of analytical solutions for the longitudinal momentum:
\begin{equation}
 p_z^{(\pm)}\equiv\frac{k_z}{2}\left[1+\frac{2(n-n^{\prime})|qB|}{\Omega^2-k_z^2}\pm\frac{\Omega}{|k_z|}\sqrt{\left(1-\frac{k_{-}^2}{\Omega^2-k_z^2}\right)\left(1-\frac{k_{+}^2}{\Omega^2-k_z^2}\right)}\right].
 \label{eq.SolutionsPz}
\end{equation}
Note that these are exactly the same as in the case of dilepton emission \cite{Wang:2022jxx}, provided the dilepton invariant mass is replaced with the scalar boson mass. Nevertheless, as we will see below, the rate and the angular distribution of scalar emission will be very different. 

By using the analytical solutions in Eq.~(\ref{eq.SolutionsPz}), we can also obtain the corresponding fermion Landau-level energies, 
\begin{eqnarray}
  \label{eq.EnergySolution1}
 E_{n,p_z}\Big|_{p_z^{(\pm)}}&=&-\frac{\eta\Omega}{2}\left[1+\frac{2(n-n^{\prime})|qB|}{\Omega^2-k_z^2}\pm\frac{|k_z|}{\Omega}\sqrt{\left(1-\frac{k_{-}^2}{\Omega^2-k_z^2}\right)\left(1-\frac{k_{+}^2}{\Omega^2-k_z^2}\right)}\right],\\
 E_{n^{\prime},p_z-kz}\Big|_{p_z^{(\pm)}}&=&\frac{\lambda\eta\Omega}{2}\left[1-\frac{2(n-n^{\prime})|qB|}{\Omega^2-k_z^2}\mp\frac{|k_z|}{\Omega}\sqrt{\left(1-\frac{k_{-}^2}{\Omega^2-k_z^2}\right)\left(1-\frac{k_{+}^2}{\Omega^2-k_z^2}\right)}\right].
  \label{eq.EnergySolution2}
\end{eqnarray}
Having explicit analytical solutions for the longitudinal momentum, now we can rewrite the Dirac $\delta$-function in Eq.~(\ref{eq.MF-VEdiagA12}) as follows:
\begin{equation}
 \delta\left(E_{n,p_z}-\lambda E_{n^{\prime},p_z-k_z}+\eta\Omega\right)=\sum_{s=\pm} \frac{2E_{n,p_z}E_{n^{\prime},p_z-k_z}}{\sqrt{\left(\Omega^2-k_z^2-k_{-}^2\right)\left(\Omega^2-k_z^2-k_{+}^2\right)}}\delta\left(p_z-p_z^{(s)}\right).
 \label{eq.DiracDeltaSimplify}
\end{equation}
Finally, by integrating over $p_z$ in Eq.~(\ref{eq.MF-VEdiagA12}), we derive the expression for the imaginary part of the scalar boson self-energy in the form of a convergent series over Landau levels:
\begin{eqnarray}
  \mbox{Im}\left[\Sigma^R(\Omega,\mathbf{k})\right]
  &=&\frac{g^2}{2\pi\ell^2}\sum_{n>n^{\prime}}^\infty\frac{\theta\left(\Omega^2-k_z^2-k_{+}^2\right)-\theta\left(k_{-}^2+k_z^2-\Omega^2\right)}{\sqrt{\left(\Omega^2-k_z^2-k_{-}^2\right)\left(\Omega^2-k_z^2-k_{+}^2\right)}}h\left(n,n^{\prime}\right)\nonumber\\
  &&\times\bigg[\left(\left(n+n^{\prime}\right)|qB|-\frac{1}{2}\left(\Omega^2-k_z^2\right)+2m^2\right)\left(\mathcal{I}^{n,n^{\prime}}_0(\xi)+\mathcal{I}^{n-1,n^{\prime}-1}_0(\xi)\right)-\frac{2}{\ell^2}\mathcal{I}^{n-1,n^{\prime}-1}_2(\xi)\bigg]\nonumber\\
  &+&\frac{g^2}{4\pi\ell^2}\sum_{n=0}^\infty\frac{\theta\left(\Omega^2-k_z^2-4m^2 -8n|qB|\right)}{\sqrt{\left(\Omega^2-k_z^2\right)\left(\Omega^2-k_z^2-4m^2 -8n|qB|\right)}}h_0\left(n\right)\nonumber\\
  &&\times\bigg[\left(2n|qB|-\frac{1}{2}\left(\Omega^2-k_z^2\right)+2m^2\right)\left(\mathcal{I}^{n,n}_0(\xi)+\mathcal{I}^{n-1,n-1}_0(\xi)\right)-\frac{2}{\ell^2}\mathcal{I}^{n-1,n-1}_2(\xi)\bigg].
 \label{eq.MF-VEdiagA15}
\end{eqnarray}
Here we introduced the following functions made of the Fermi-Dirac distributions:
\begin{eqnarray}
  \label{eq.hfunction1}
 h\left(n,n^{\prime}\right)&\equiv& 2-\sum_{s_1,s_2=\pm}n_F\left(\frac{\Omega}{2}+s_1\frac{\Omega\left(n-n^{\prime}\right)|qB|}{\Omega^2-k_z^2}+s_2\frac{|k_z|}{2\left(\Omega^2-k_z^2\right)}\sqrt{\left(\Omega^2-k_z^2-k_{-}^2\right)\left(\Omega^2-k_z^2-k_{+}^2\right)}\right),\\
 h_0\left(n\right)&\equiv& h(n,n)=2-2\sum_{s_2=\pm}n_F\left(\frac{\Omega}{2}+s_2\frac{|k_z|}{2}\sqrt{1-\frac{4\left(m^2+2n|qB|\right)}{\Omega^2-k_z^2}}\right).
  \label{eq.hfunction2}
\end{eqnarray}
Notice that the second term in Eq.~(\ref{eq.MF-VEdiagA15}) is the contribution due to annihilation processes with $n=n^{\prime}$.

The expression for the imaginary part of self-energy (\ref{eq.MF-VEdiagA15}) is one of the main analytical results of this study. By substituting it into the definition in Eq.~(\ref{eq:diff-rate}), we can calculate the differential emission rate of neutral bosons from a magnetized plasma. The corresponding numerical results will be presented and analyzed in the next section. 

Note that the general structure of the expression in Eq.~(\ref{eq.MF-VEdiagA15}) resembles the photon polarization tensor obtained in Ref.~\cite{Wang:2021ebh}. However, there are some profound differences. Unlike spin-one photons, the bosons are spin-zero particles in the model at hand. As we discuss later in detail, the spinless nature of bosons strongly affects the angular dependence of the self-energy and, in turn, the corresponding angular distribution of boson emission. For example, the differential rate due to particle-splitting processes will be suppressed in the direction parallel to the magnetic field. In the case of photons, such emission was not only allowed but played a dominant role at small energies. 

Before concluding this subsection, it is instructive to consider a simplified kinematic regime with $\mathbf{k}_\perp=0$  (i.e., for $\theta=0$ or $\theta=\pi$). It is the only case that was analyzed previously in the literature, see Ref.~\cite{Bandyopadhyay:2018gbw}. It corresponds to scalar boson emission in the direction of the magnetic field. Substituting $\mathbf{k}_\perp=0$, the general result for self-energy in Eq.~(\ref{eq.MF-VEdiagA15}) reduces down to
\begin{equation}
  \mbox{Im}\left[\Sigma^R(\Omega,\mathbf{0},k_z)\right]
  =-\frac{g^2}{8\pi\ell^2} \frac{\left(\Omega^2-k_z^2-4m^2\right)}{\sqrt{\Omega^2-k_z^2}} 
  \sum_{n=0}^\infty \alpha_n  \frac{\theta\left(\Omega^2-k_z^2-4m^2 -8n|qB|\right) }{\sqrt{\Omega^2-k_z^2-4m^2 -8n|qB|}} h_0\left(n\right)  ,
 \label{eq.MF-VEdiagA16}
\end{equation}
where we introduced $\alpha_n = 2-\delta_{n,0}$ and took into account that $\lim_{\xi\to 0}\left[\mathcal{I}^{n,n^{\prime}}_0(\xi) \right] =\delta_{n,n^{\prime}}$ and $\lim_{\xi\to 0}\left[\mathcal{I}^{n,n^{\prime}}_2(\xi)\right] =2(n+1)\delta_{n,n^{\prime}}$ \cite{Wang:2021ebh}. Compared to the general result in Eq.~(\ref{eq.MF-VEdiagA15}), this expression for the self-energy is much simpler. More importantly, from a physics viewpoint, the kinematics of allowed processes is very restrictive at $\mathbf{k}_\perp=0$. In particular, no one-to-two particle-splitting processes contribute in this case at all. Only the particle-antiparticle annihilation processes do (and only if $M>2m$). Since the same does not hold at any nonzero $\mathbf{k}_\perp$, such a simplified regime is an exceptional outlier. Furthermore, as we will see in the next section, the particle-splitting processes contribute substantially to the total emission rate in some kinematic regimes.

\subsection{Zero magnetic field limit}
\label{subsec:B0case}

 Here we verify that the result for the self-energy in Eq.~(\ref{eq.MF-VEdiagA15}) is consistent with the known zero-field limit. For our purposes, it is sufficient to consider only the case with $\mathbf{k}_\perp=0$. 

To consider the limit of vanishing magnetic field in Eq.~(\ref{eq.MF-VEdiagA16}), we introduce a continuous variable $v =2n|qB|$ and replace the sum over $n$ with an integral over $v$. Then, we have
\begin{equation}
\mbox{Im}\left[\Sigma^R(\Omega,\mathbf{k})\right]
  =-\frac{g^2}{8\pi} \frac{\left(\Omega^2-|\mathbf{k}|^2-4m^2\right)}{\sqrt{\Omega^2-|\mathbf{k}|^2}} \theta\left(\Omega^2-|\mathbf{k}|^2-4m^2\right)
  \int_{0}^{v_0} \frac{dv}{\sqrt{v_0-v}} \left[1-\sum_{s_2=\pm}n_F\left(\frac{\Omega}{2}+s_2\frac{|\mathbf{k}|\sqrt{v_0-v}}{\sqrt{\Omega^2-|\mathbf{k}|^2}}\right)\right],
 \label{eq.MF-VEdiagA17}
\end{equation}
where the upper limit of integration is $v_0=\left(\Omega^2-|\mathbf{k}|^2-4m^2\right)/4$. In the last expression, we also replaced $|k_z|$ with $|\mathbf{k}|$ in view of the Lorentz symmetry, which is restored in the absence of a magnetic field. 

After introducing the new integration variable $u=|\mathbf{k}|\sqrt{v_0-v}/\sqrt{\Omega^2-|\mathbf{k}|^2}$, we obtain
\begin{equation}
\mbox{Im}\left[\Sigma^R(\Omega,\mathbf{k})\right]
  =-\frac{g^2}{4\pi} \frac{\left(\Omega^2-k_z^2-4m^2\right)}{|k_z|} 
  \theta\left(\Omega^2-k_z^2-4m^2\right)
  \int_{0}^{u_0} du \left[1-\sum_{s_2=\pm}n_F\left(\frac{\Omega}{2}+s_2 u \right)\right],
 \label{eq.MF-VEdiagA17a}
\end{equation}
where 
\begin{equation}
u_0 = \frac{|\mathbf{k}|\sqrt{\Omega^2-|\mathbf{k}|^2-4m^2}}{2\sqrt{\Omega^2-|\mathbf{k}|^2}}.
\end{equation}
Finally, after integrating over $u$, we derive 
\begin{equation}
\mbox{Im}\left[\Sigma^R(\Omega,\mathbf{k})\right]
  =-\frac{g^2}{8\pi} \left(\Omega^2-|\mathbf{k}|^2-4m^2\right) 
  \left[\frac{\sqrt{\Omega^2-|\mathbf{k}|^2-4m^2}}{\sqrt{\Omega^2-|\mathbf{k}|^2}} +\frac{2T}{|\mathbf{k}|}
  \ln\frac{1+e^{-E_{+}/T}}{1+e^{-E_{-}/T}}\right] \theta\left(\Omega^2-|\mathbf{k}|^2-4m^2\right).
 \label{eq.MF-VEdiagA17b}
\end{equation}
Note that $E_{\pm}\equiv \Omega/2 \pm u_0$ coincide with the definitions in Eq.~(\ref{eq.qB=0.Energies}) in Appendix~\ref{app.IntegrationB=0}. The final result for the imaginary part of self-energy in Eq.~(\ref{eq.MF-VEdiagA17b}) also agrees with the $B=0$ expression given in Eq.~(\ref{eq.qB=0.self-energy7}).

When the scalar bosons are on the mass shell, i.e., $\Omega^2=M^2+|\mathbf{k}|^2$, one has 
\begin{eqnarray}
\mbox{Im}\left[\Sigma^R (\mathbf{k})\right]\Big|_{\Omega^2=M^2+|\mathbf{k}|^2}=-\frac{g^2}{8\pi} \left(M^2-4m^2\right) 
  \left[\frac{\sqrt{M^2-4m^2}}{\sqrt{M^2}} +\frac{2T}{|\mathbf{k}|}
  \ln\frac{1+e^{-E_{+}/T}}{1+e^{-E_{-}/T}}\right] \theta\left(M^2-4m^2\right).
 \label{eq.MF-VEdiagA18}
\end{eqnarray}
As we see, this expression is nonvanishing only when $M^2\geq 4m^2$. From a physics viewpoint, it indicates that the annihilation processes are the only ones contributing. It is not surprising since one-to-two particle-spitting processes ($\psi\rightarrow \psi+\phi $ and $\bar{\psi} \rightarrow \bar{\psi}+\phi $) are forbidden without a background magnetic field. The latter is evident when considering the process in the rest frame of the boson. (Curiously, such one-to-two processes may be allowed when the masses of the initial and final fermions are different \cite{Bastero-Gil:2010dgy}.) In the case of a nonzero magnetic field, in contrast, particle-spitting processes are allowed because the momentum conservation constraint in the plane perpendicular to the field is relaxed.

\section{Numerical results}
\label{sec.num}

Here, we use the imaginary part of self-energy derived in the previous section to analyze the differential emission rate of neutral bosons from a magnetized plasma. Because of an elaborate expression in Eq.~(\ref{eq.MF-VEdiagA15}) and the  complications due to the sum over Landau levels, the angular dependence of the rate in Eq.~(\ref{eq:diff-rate}) is hard to comprehend. Therefore, here we study it with the help of numerical methods. 

In the model at hand, two qualitatively different regimes exist. They are determined by the value of the scalar boson mass $M$, which can be either greater or less than the fermion-antifermion threshold $2m$. In the subthreshold regime ($M<2m$), 
no scalar boson production can occur without a background magnetic field at the leading order in coupling. The situation changes when $B\neq 0$. The annihilation becomes possible when the scalar boson energy exceeds the threshold of $2m$. More interestingly, the boson production via particle-splitting processes is allowed in the whole range of energies $\Omega>M$.

Below, we will study both regimes by considering the following two representative cases: $M=3m$ (suprathreshold) and $M=m/3$ (subthreshold). In each case, we will study the angular dependence of the rate in detail for several representative values of the magnetic field and temperature. As we will see, the behavior of the differential rates will be very different, especially at small values of the polar angle $\theta$.  

To reduce the number of free parameters and simplify the analysis, we will express all dimensionful quantities in units of the fermion mass $m$. We will consider two different values of the magnetic field, i.e., $|qB|=(2m)^2$ (moderate field) and $|qB|=(5m)^2$ (strong field), and two different temperatures, i.e., $T=5m$ and $T=15m$, that correspond to moderately relativistic and ultrarelativistic plasmas, respectively. Without loss of generality, we will use the Yukawa coupling $g=1$ in numerical calculations below.

When calculating numerically the imaginary part of self-energy (\ref{eq.MF-VEdiagA15}), one needs to sum over Landau-level indices $n$ and $n^{\prime}$. Since the corresponding double-series is convergent, one may truncate the summation at sufficiently large finite $n_{\rm max}$. Its value will be determined by the largest energy scale in the problem, which will be set by either the temperature or the scalar boson energy $\Omega$. The latter will be varied in a range from $\Omega =M$ up to about $\Omega \simeq 35 m$ (for $|qB|=4m^2$) and $\Omega \simeq 90 m$ (for $|qB|=25m^2$). Thus, from general considerations, one should include at least sufficient number of Landau levels to open the phase space for the processes with the largest energies. This leads to the bound from below: 
\begin{equation}
n_{\rm max} \gtrsim \left[ \mbox{max}\left\{\frac{T^2}{2|qB|}, \frac{\Omega^2}{2|qB|}\right\}\right], 
\end{equation}
where the square brackets represent the integer part.

\subsection{Moderate magnetic field, $\mathbf{|qB|=4m^2}$}

Let us start the study of the differential rate as a function of the angular coordinate $\theta$ in the case of a moderately strong magnetic field $|qB|=(2m)^2$. To achieve a high angular resolution, we will use the discretization steps of $\Delta\theta = \pi/(2n_\theta)$ with $n_\theta=10^3$. The direction along the magnetic field corresponds to $\theta=0$, and the perpendicular direction is $\theta=\pi/2$. There is no need to consider $\theta>\pi/2$, as the corresponding rates can be obtained using the symmetry with respect to mirror reflection in the $xy$-plane. Indeed, such a symmetry remains unbroken in the presence of a constant background magnetic field.

\begin{figure}[t]
 \includegraphics[width=0.45\textwidth]{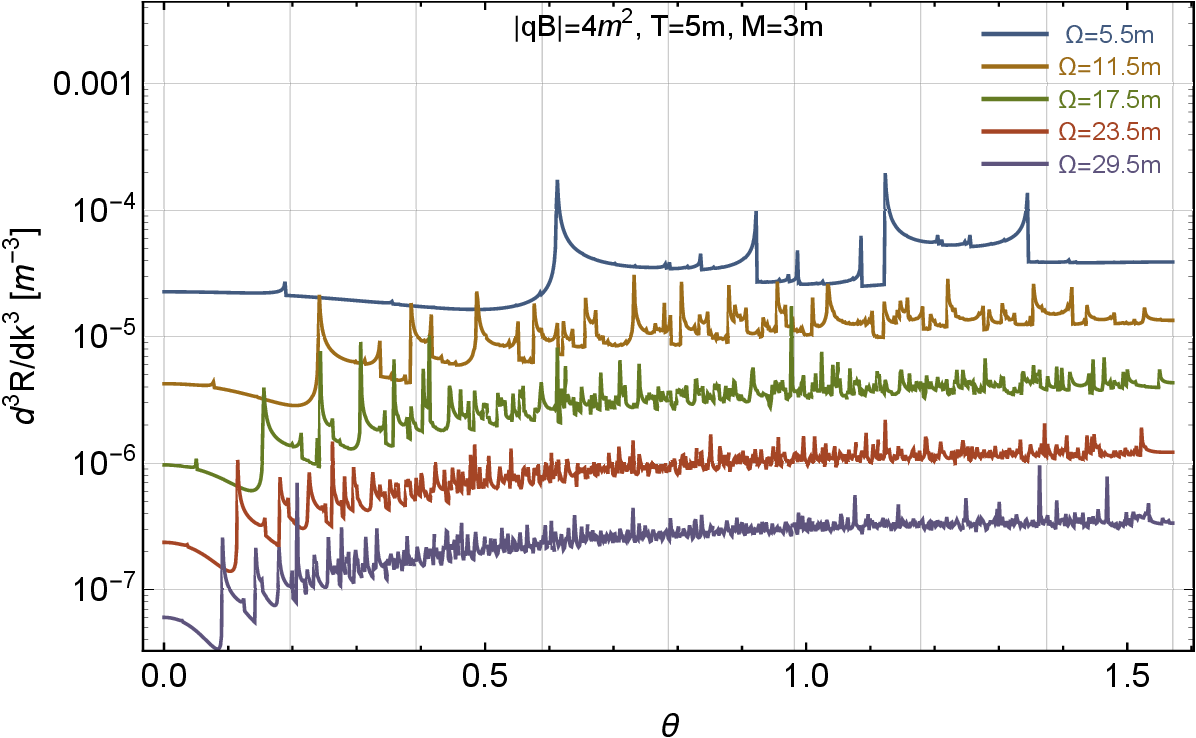}
 \hspace{0.04\textwidth}
 \includegraphics[width=0.45\textwidth]{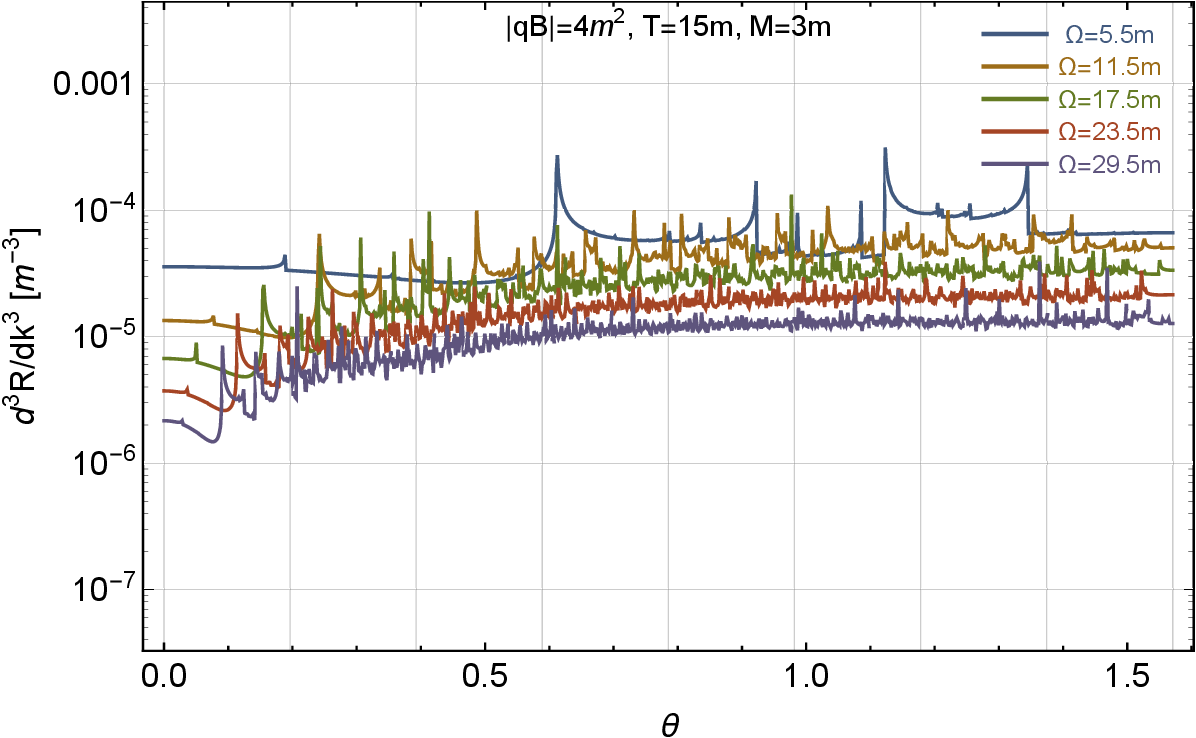}\\[2mm]
  \includegraphics[width=0.45\textwidth]{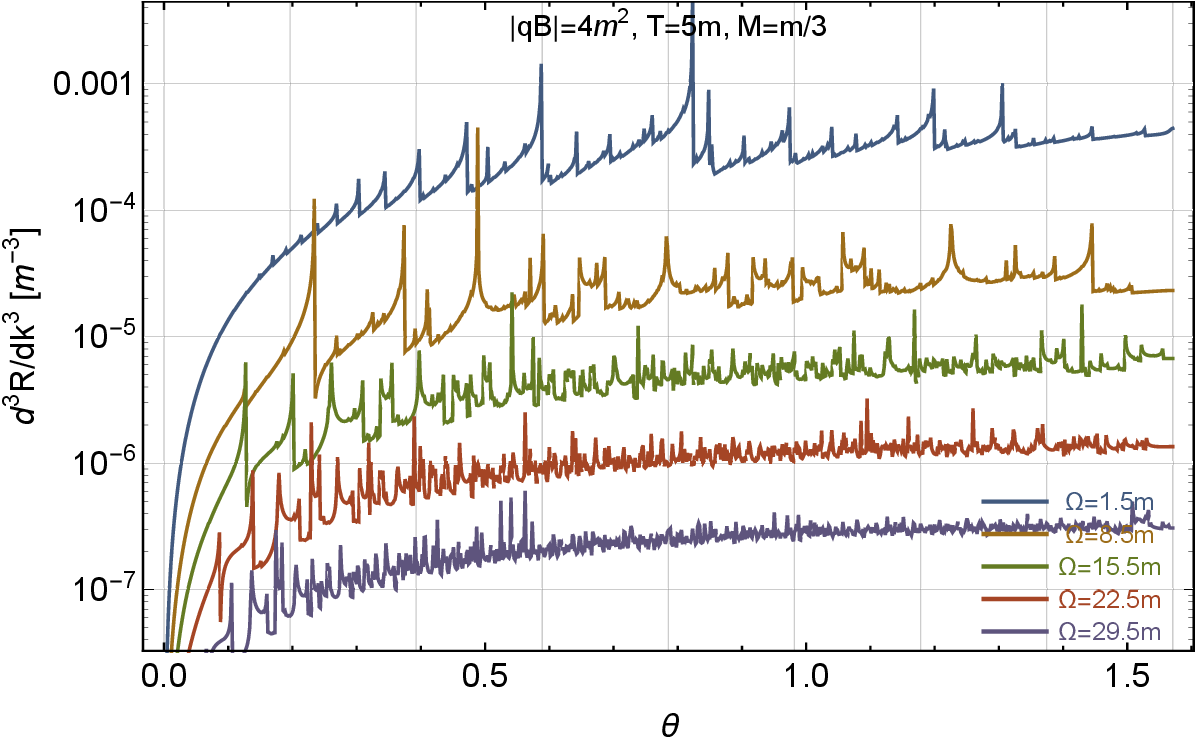}
 \hspace{0.04\textwidth}
 \includegraphics[width=0.45\textwidth]{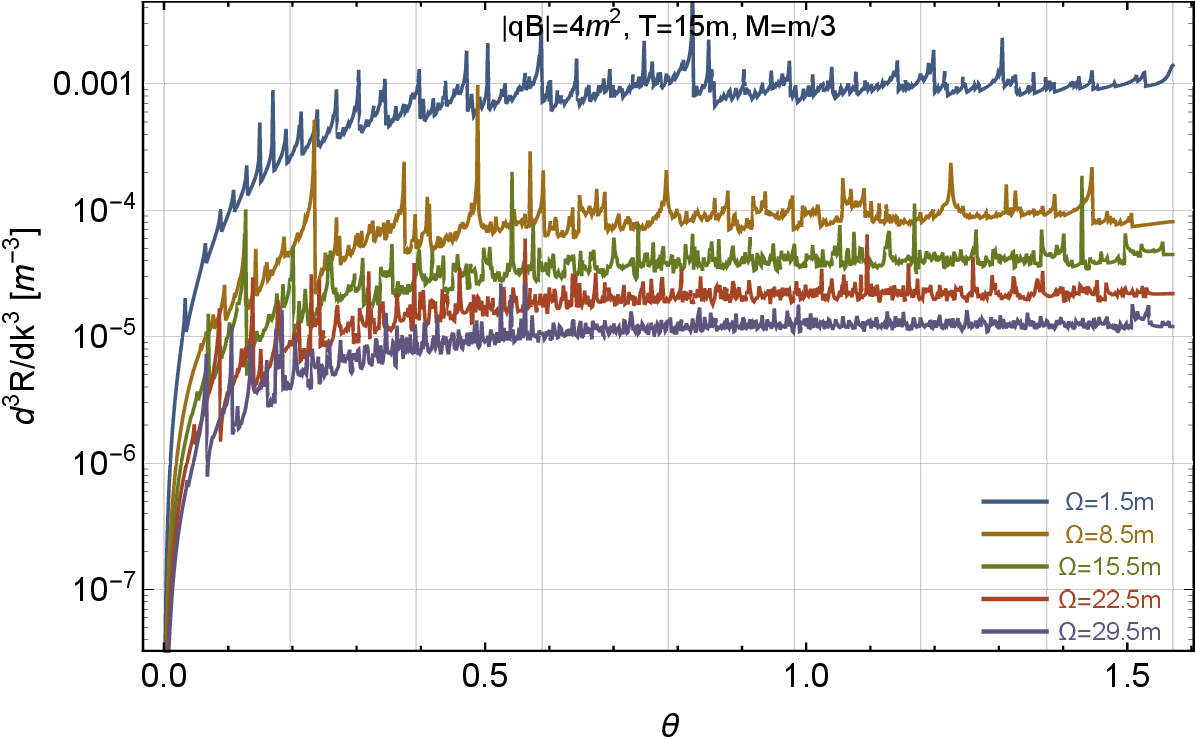}
 \caption{Neutral scalar boson differential production rates for several different energies and two fixed temperatures:  $T=5m$ (left panels) and $T=15m$ (right panels). The magnetic field is $|qB|=4m^2$ and the scalar boson masses are $M=3m$ (top panels) and $M=m/3$ (bottom panels).}
 \label{fig:M3_B4_OmegaS}
\end{figure}

Representative numerical results for the differential rates are shown in Fig.~\ref{fig:M3_B4_OmegaS} for two fixed temperatures, i.e., $T=5m$ (left panels) and $T=15m$ (right panels), as well as two choices of the scalar boson mass, i.e., $M=3m$ (top panels) and $M=m/3$ (bottom panels). 
Different lines correspond to different energies of neutral scalar bosons. They satisfy the mass-shell condition: $\Omega =\sqrt{M^2+k_\perp^2+k_z^2}$, where $k_\perp = k\sin{\theta}$ and $k_z = k\cos{\theta}$.

By comparing the results for two different temperatures in the left and right panels of Fig.~\ref{fig:M3_B4_OmegaS}, we see that the rates tend to grow with temperature, as expected. In the case of $M=3m$, the growth is relatively week at first when the energy exceeds the threshold $\Omega\gtrsim M$ only slightly. It becomes more pronounced at higher values of energy. From a different perspective, the average rates decrease with increasing the scalar boson energy. However, one sees a substantial suppression only after the energy exceeds the plasma's temperature. To a large degree, such a behavior is dominated by the annihilation processes. 

It is worth noting that the contribution of the lowest Landau level to the total rate remains relatively modest across the whole range of scalar boson energies. It plays a significant role only in the suprathreshold case ($M=3m$) at small temperatures, when the scalar boson's energy is only slightly higher than its minimum value $\Omega_{\rm min}= M$. This observation underscores the limitations of the so-called lowest Landau level approximation, which is often employed to obtain simple estimates in the strong field regime. As we see, in relativistic plasmas, relying on such an approximation would yield unreliable results.

The growth of rates with increasing temperature is more pronounced in the case of a subthreshold scalar boson mass, i.e., $M=m/3$, as seen from the two bottom panels of Fig.~\ref{fig:M3_B4_OmegaS}. The qualitative behavior is also different, especially at small values of the polar angle $\theta$. To understand this subthreshold regime, it is important to remember that the scalar production is possible only because of a nonzero magnetic field. Since $M<2m$, neither annihilation nor (anti)particle-splitting processes can occur at $\theta=0$, see Eq.~(\ref{eq.MF-VEdiagA16}) and related discussion. This is in drastic difference to the suprathreshold case in the two top panels of Fig.~\ref{fig:M3_B4_OmegaS}.

\begin{figure}[t]
 \includegraphics[width=0.31\textwidth]{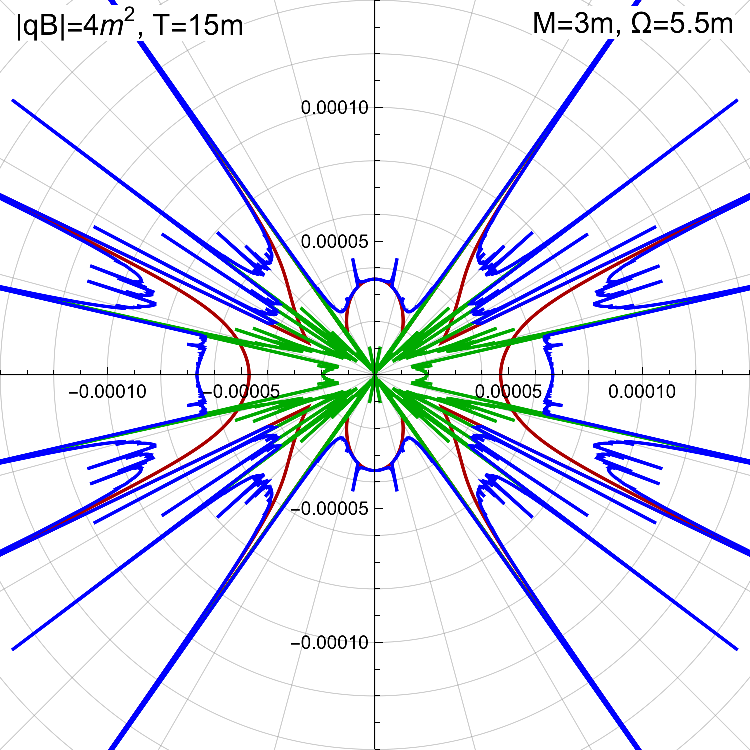}
 \hspace{0.02\textwidth}
 \includegraphics[width=0.31\textwidth]{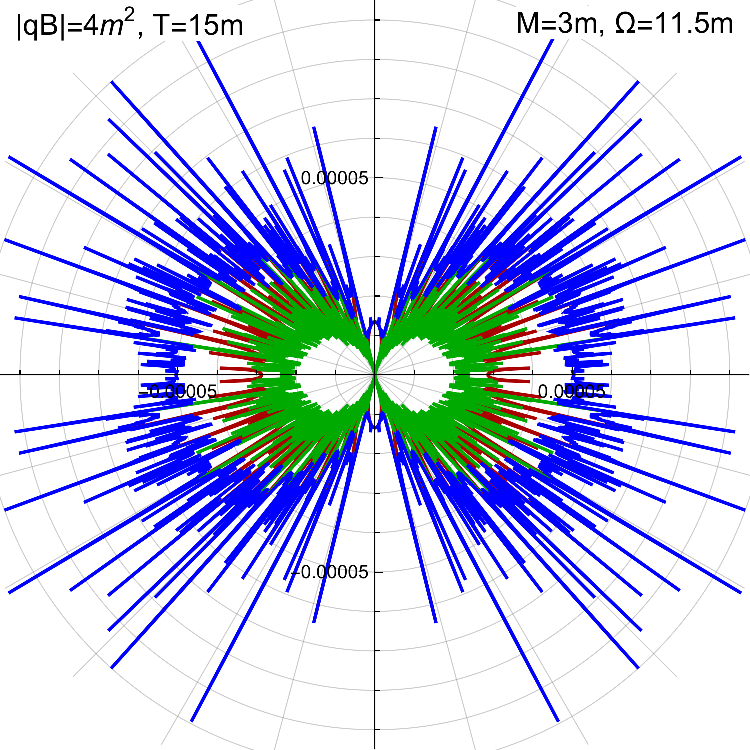}
 \hspace{0.02\textwidth}
  \includegraphics[width=0.31\textwidth]{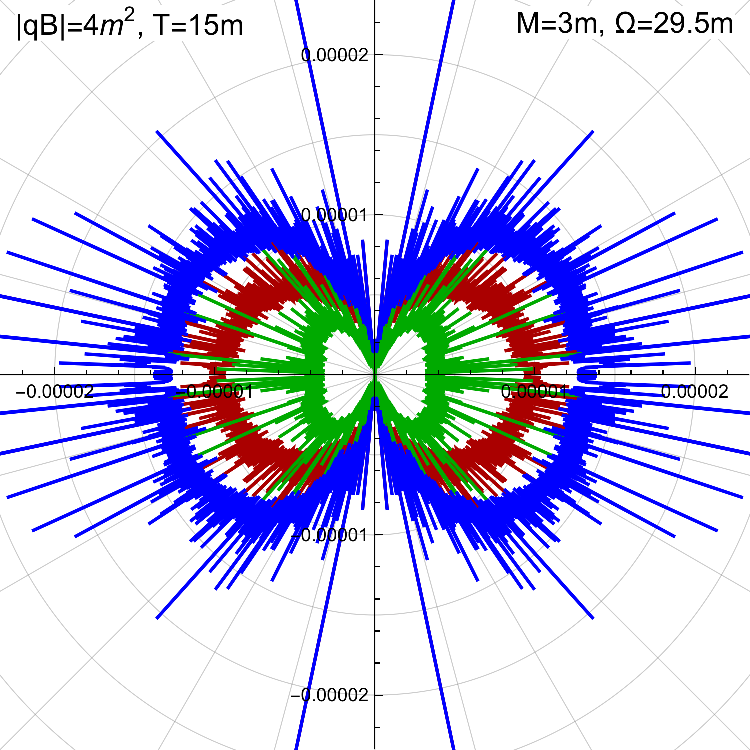}\\[2mm]
 \includegraphics[width=0.31\textwidth]{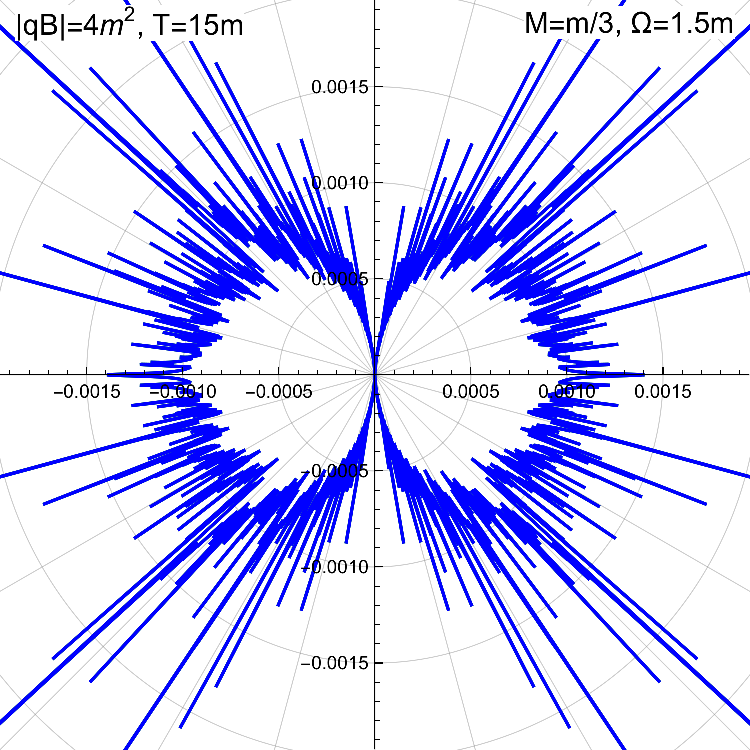}
 \hspace{0.02\textwidth}
 \includegraphics[width=0.31\textwidth]{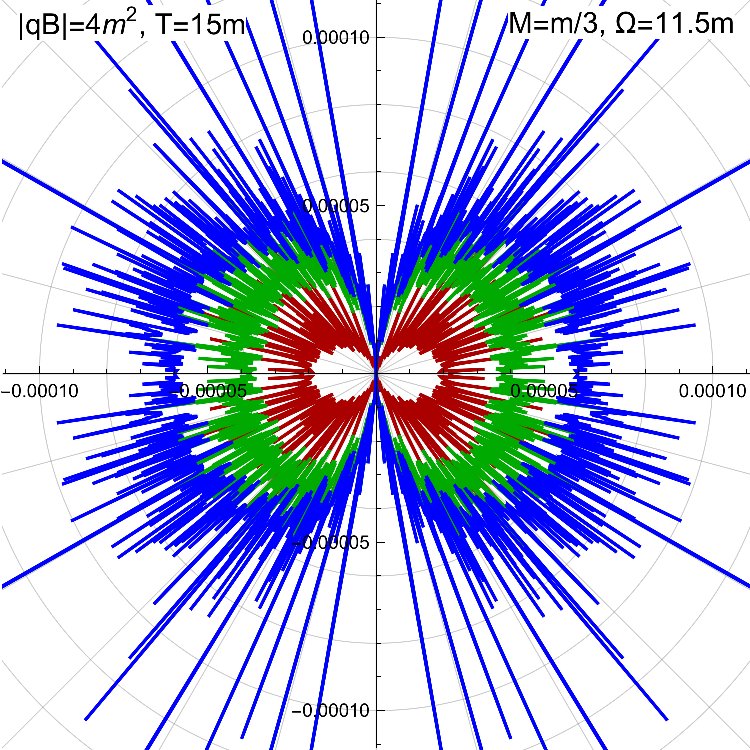}
 \hspace{0.02\textwidth}
  \includegraphics[width=0.31\textwidth]{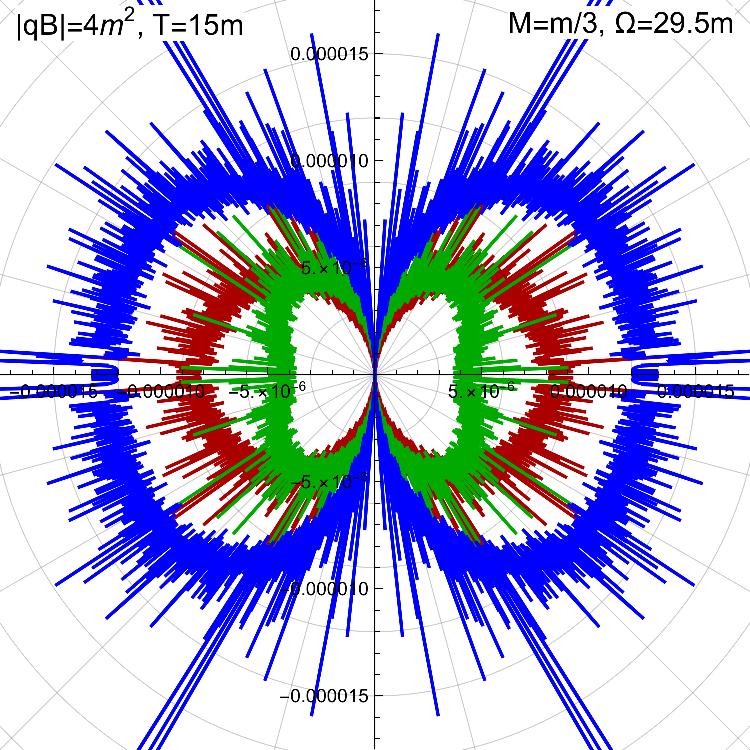}
 \caption{The angular profile of the scalar boson production rates for several different energies and fixed temperature $T=15m$. The magnetic field is $|qB|=4m^2$ and the scalar boson masses are $M=3m$ (top panels) and $M=m/3$ (bottom panels). Each panel contains separate contributions due to annihilation (red lines) and particle-splitting (green lines) processes, as well as the total rates (blue lines).}
 \label{fig:polar_B4_OmegaS}
\end{figure}

For both temperatures and both values of the scalar mass, the differential rates tend to grow on average as a function of $\theta$. It implies that the scalar bosons are emitted predominantly in the directions perpendicular to the magnetic field. We can easily visualize the corresponding emission profiles using the polar plots in Fig.~\ref{fig:polar_B4_OmegaS}. According to our convention, the magnetic field points upwards. The six individual panels show the polar plots for emission rates of bosons with several fixed energies and the two mass choices: $M=3m$ (top panels) and $M=m/3$ (bottom panels). The red lines represent the partial contributions of the annihilation rates, the green lines represent the particle-splitting rates, and the blue lines give the total rates. We show the results only for one temperature, $T=15m$. The results for another temperature ($T=5m$) are qualitatively similar but have different magnitudes and contain slightly different admixture of annihilation and particle-splitting processes. Their relative contributions will become clear when we discuss the total rates below.

As seen from Fig.~\ref{fig:polar_B4_OmegaS}, both annihilation (red lines) and particle-splitting (green lines) processes tend to provide higher rates of the scalar boson production in the directions perpendicular to the magnetic field. While having similar butterfly-shaped profiles, relative magnitudes of the two types of contributions vary with model parameters. In the suprathreshold case $M=3m$, annihilation dominates almost at all energies. In the subthreshold case $M=m/3$, however, the particle-splitting processes contribute more at small energies, but annihilation overtakes them at large energies. It is interesting to draw attention to the spacial case of $\Omega=1.5m$ when the boson mass is $M=m/3$, which falls into the subthreshold regime with $M<\Omega<2m$. In this case, particle-splitting processes are the only ones contributing to the total rate. It is the reason why the corresponding (bottom left) panel in Fig.~\ref{fig:polar_B4_OmegaS} has only blue lines visible. (Technically, the green lines, with the exact same profile, hide behind the blue ones.)

Let us now turn to the total rate $dR/d\Omega$ integrated over all angular directions, as defined in Eq.~(\ref{eq:diff-rate.dOmega}). It describes production rate (per unit time and unit volume) of scalar bosons with energies between $\Omega$ and $\Omega+d\Omega$. Unlike the differential rate, its expression contains an extra power of momentum, which accounts for the available phase space. Clearly, such a phase space collapses when $\Omega$ approaches $M$ from above. Then, the rate $dR/d\Omega$ should also vanish when $\Omega\to M$. We will see below that it is indeed the case. The extra power of the momentum in the definition will also explain why $dR/d\Omega$ does not start decreasing with energy until $\Omega$ becomes several times the plasma's temperature. 

\begin{figure}[t]
\includegraphics[width=0.45\textwidth]{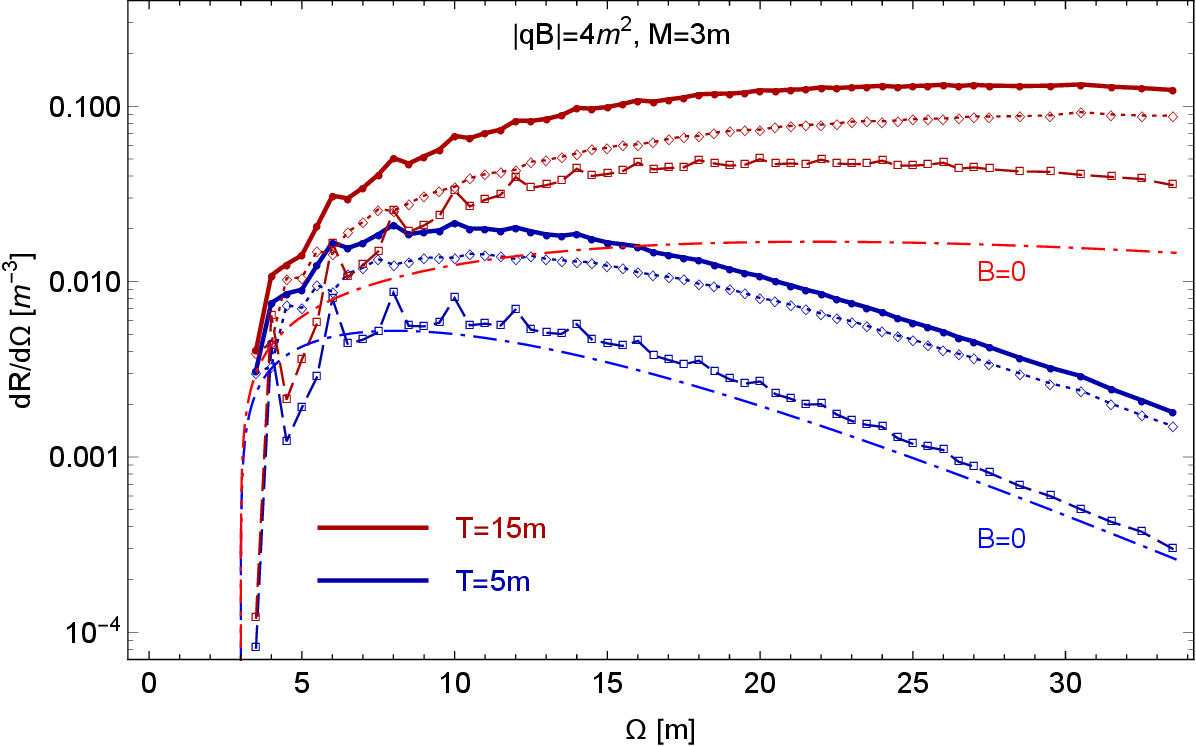}
\hspace{0.04\textwidth}
\includegraphics[width=0.45\textwidth]{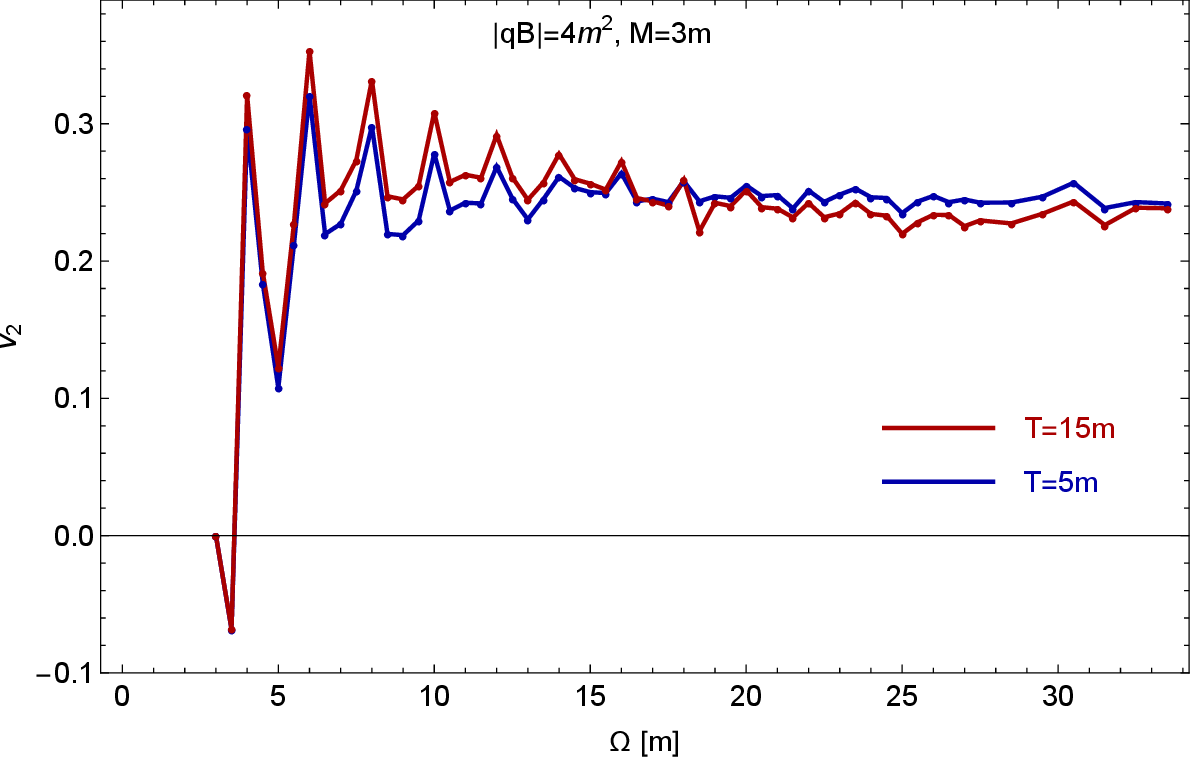}\\[2mm]
\includegraphics[width=0.45\textwidth]{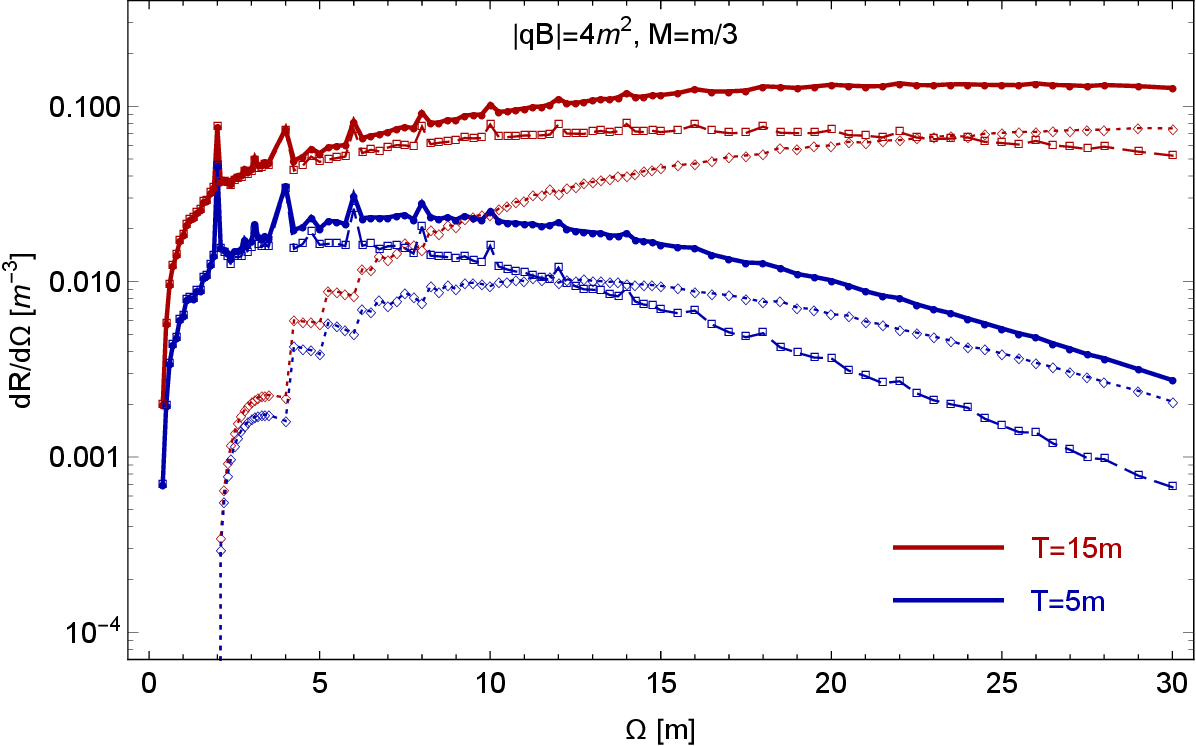}
\hspace{0.04\textwidth}
\includegraphics[width=0.45\textwidth]{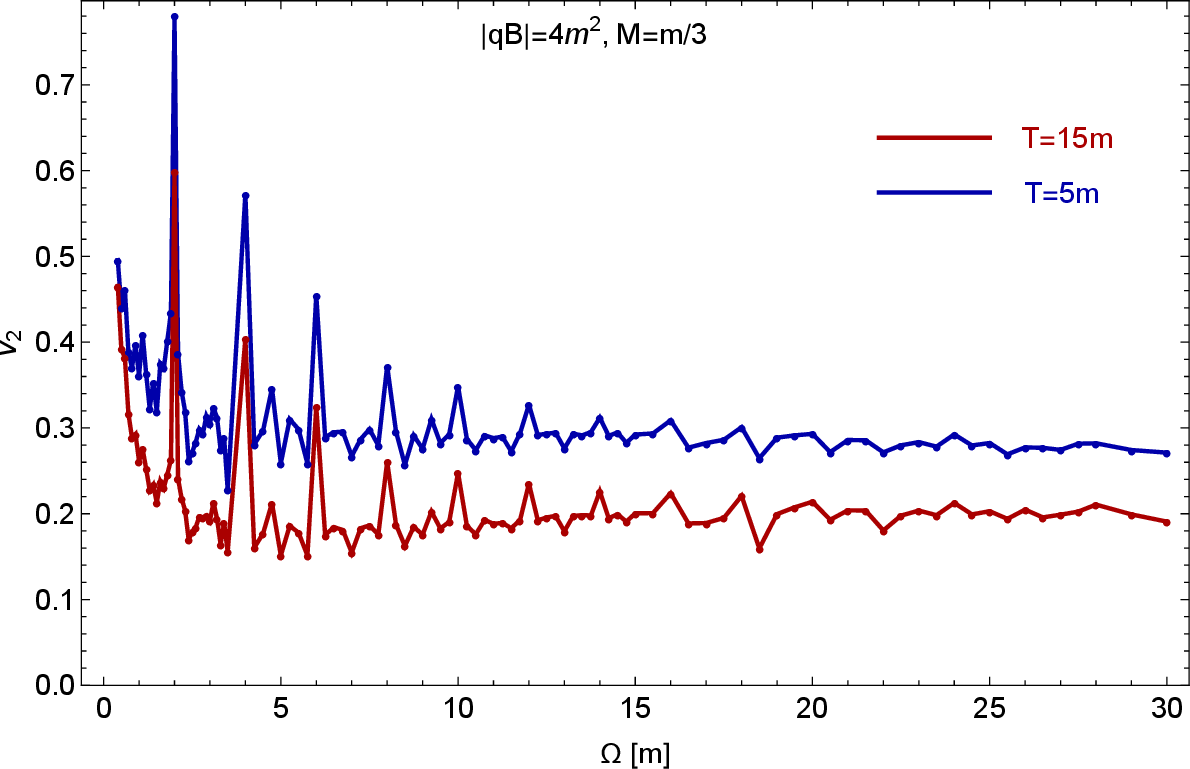}
\caption{The total rates and ellipticity of scalar boson emission from a magnetized plasma at two different temperatures: $T=5m$ (blue lines) and $T=15m$ (red lines). The magnetic field is $|qB|=4m^2$ and the scalar boson masses are $M=3m$ (top panels) and $M=m/3$ (bottom panels).}
\label{fig:total-ratesB4}
\end{figure}

For the case of the moderately strong field $|qB|=4m^2$, the corresponding rates as functions of the energy are summarized in the two left panels of Fig.~\ref{fig:total-ratesB4}. The other two panels on the right show the ellipticity measure $v_2$ for the scalar boson emission, formally defined by Eq.~(\ref{eq:v2}). In all the panels, the color coding represents temperature, with the blue for $T=5m$ and the red for $T=15m$. In addition to the total rates (filled circles) shown in the panels on the left, we also display the separate partial contributions due to annihilation (open diamonds) and particle-splitting (open squares) processes. For guiding the eye, we connected the points with different lines: solid (total rate), dotted (annihilation part) and dashed (particle-splitting part), respectively. For comparison, the dash-dotted lines represent the rates in the limit of the vanishing magnetic field. As we argued before, such a limit is meaningful only for $M=3m$ (suprathreshold case). For subthreshold scalar mass $M=m/3$, the rates vanish without a magnetic field. 

The rates for all model parameters represented in Fig.~\ref{fig:total-ratesB4} share many similar features. Overall, they have a tendency to grow with increasing the temperature. It is easy to understand since the number densities of both occupied positive-energy states and unoccupied negative-energy states increase with temperature. The availability (anti)particles in such states, in turn, opens the phase space for all relevant processes producing scalar bosons. On the other hand, as a function of energy, the rates grow at first, reach a maximum value around $\Omega \sim 1.7 T$, and then decrease. After passing the peak value, the behavior at high energies gradually approaches an exponential asymptote, i.e., $dR/d\Omega \sim \exp\left(-\Omega/T\right)$. 

By comparing the partial contributions of different types of processes in the two left panels of Fig.~\ref{fig:total-ratesB4}, we see that it is the annihilations rather than the particle splittings that dominate at sufficiently large energies. The interplay of the two is more subtle at low energies, where the relative contributions depend on the scalar boson mass. For the suprathreshold mass, $M = 3m$, the annihilation is more likely to dominate the total rate for (almost) all energies. For the subthreshold mass, $M = m/3$, on the other hand, the particle-splitting processes give larger contributions in a range of small energies, $\Omega \lesssim 1.7 T$. Still, even for $M = m/3$, the annihilation eventually takes over at higher energies. 

Now let us turn to the results for the ellipticity parameter $v_2$, shown in the two right panels of Fig.~\ref{fig:total-ratesB4}. In general, as we see, the values of $v_2$ are positive and relatively large. At high energies, typical values of $v_2$ are of the order of $0.2$ to $0.3$. The values tend to go down with increasing the temperature, though. There are some qualitative differences between the cases of $M = 3m$ (suprathreshold) and $M = m/3$ (subthreshold), especially in the range of small energies, i.e., $\Omega \lesssim 1.7 T$. In particular, for $M = 3m$, the ellipticity parameter $v_2$ shows a wide range of variations at small energies. It can even take negative values. These variations come from a finite number of quantum transitions between Landau levels that produce large threshold spikes in some directions and, thus, dramatically affecting $v_2$. In contrast, for $M = m/3$, the ellipticity parameter tends to grow by a factor of two or more with decreasing the energy from $\Omega =2m$ down to $\Omega =m/3$. Recall that, in this energy range, many particle-splitting processes contribute. They do not allow scalar boson production in the direction $\theta=0$ and, thus, tend to give large $v_2$.

\subsection{Strong magnetic field, $\mathbf{|qB|=25m^2}$}

\begin{figure}[t]
 \includegraphics[width=0.45\textwidth]{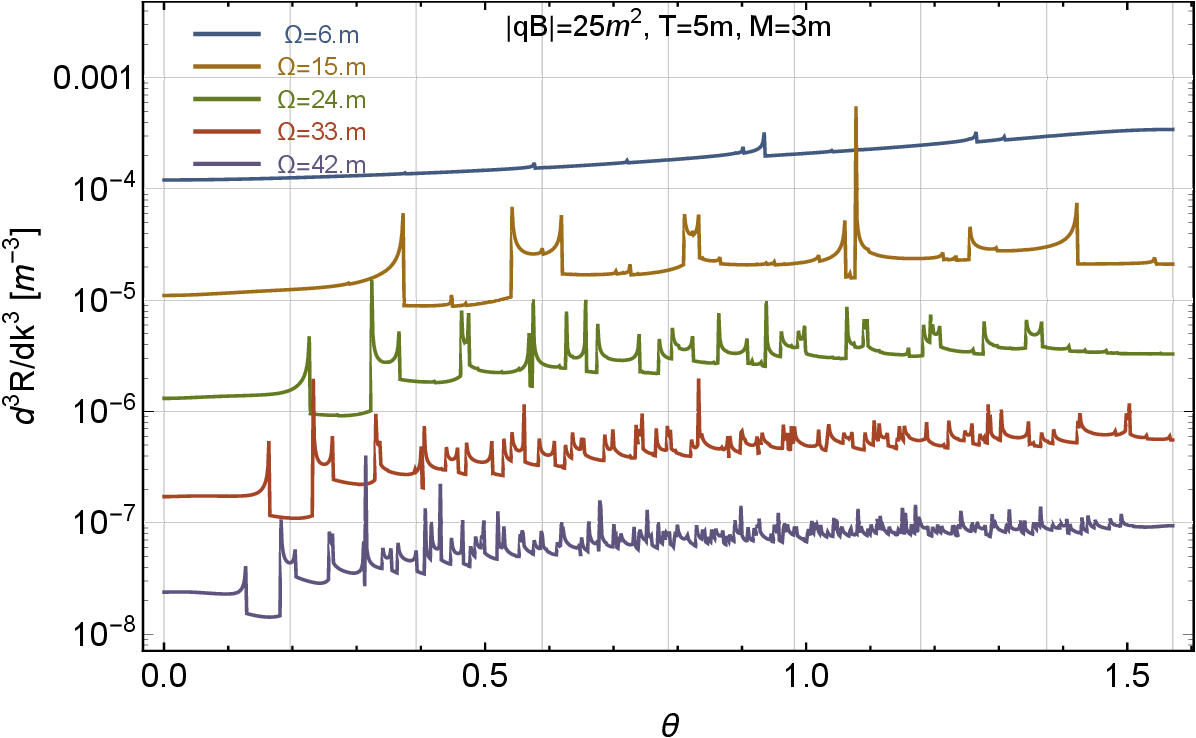}
 \hspace{0.04\textwidth}
 \includegraphics[width=0.45\textwidth]{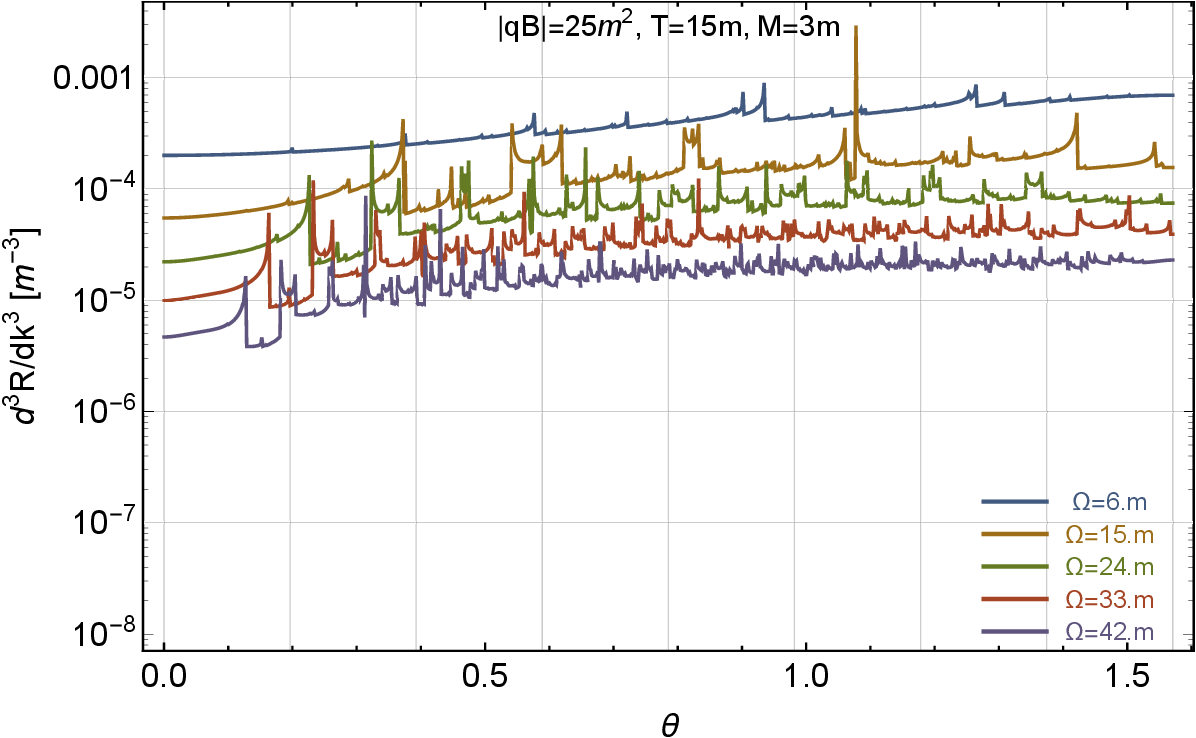}\\[2mm]
 \includegraphics[width=0.45\textwidth]{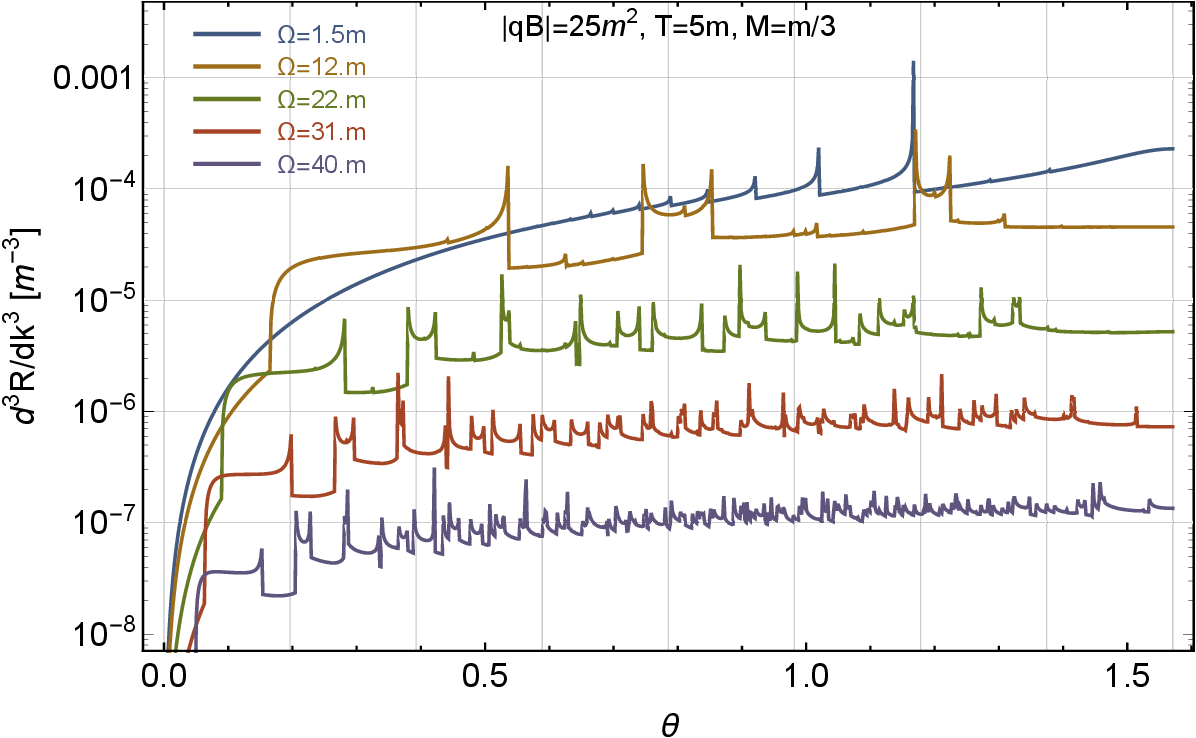}
 \hspace{0.04\textwidth}
 \includegraphics[width=0.45\textwidth]{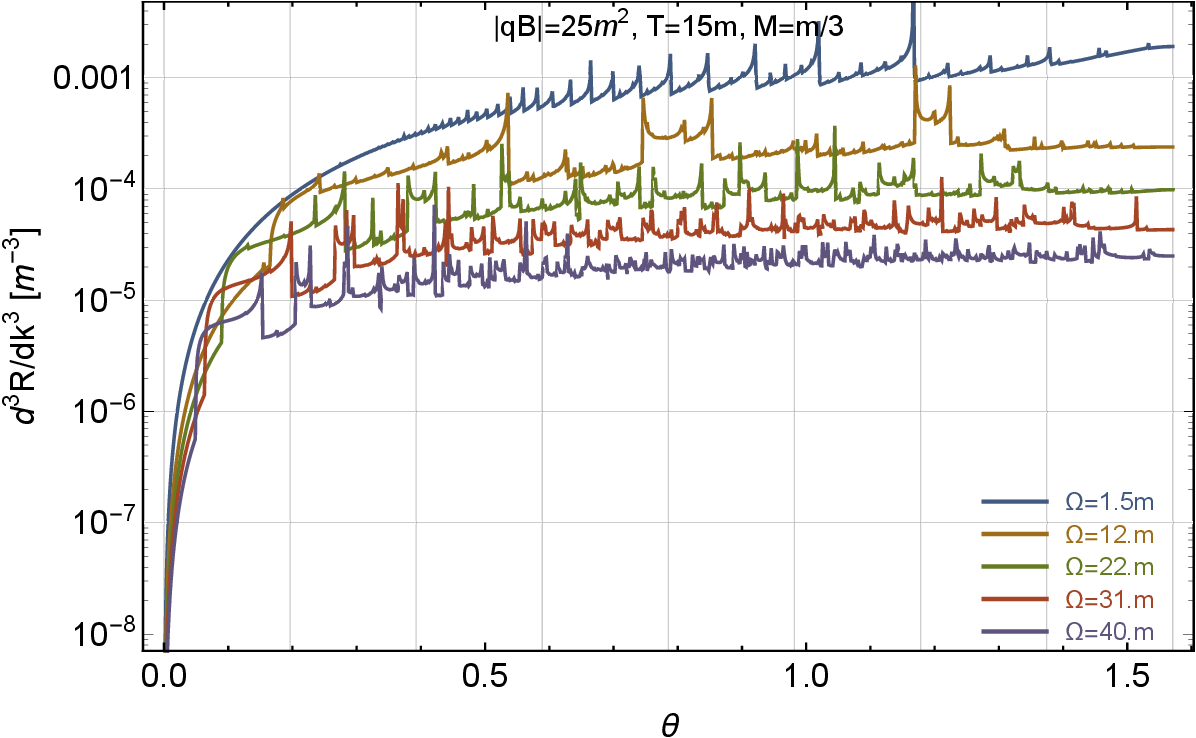}
 \caption{Neutral scalar boson differential production rates for several different energies and two fixed temperatures:  $T=5m$ (left panels) and $T=15m$ (right panels). The magnetic field is $|qB|=25m^2$ and the scalar boson masses are $M=3m$ (top panels) and $M=m/3$ (bottom panels).}
 \label{fig:M3_B25_OmegaS}
\end{figure}

Now let us consider the case of a strong field, i.e., $|qB|=(5m)^2$. As in the previous subsection, we will start from the representative numerical results for the differential rates as functions of the angular coordinate $\theta$. The rates for several fixed values of the scalar boson energy are displayed in Fig.~\ref{fig:M3_B25_OmegaS}. It includes four panels with the results for two different temperatures,  $T=5m$ (left panels) and $T=15m$ (right panels), and two different scalar boson masses, $M=3m$ (top panels) and $M=m/3$ (bottom panels).

The strong field results in Fig.~\ref{fig:M3_B25_OmegaS} are qualitatively similar to those in the weaker field, presented earlier in  Fig.~\ref{fig:M3_B4_OmegaS}. As before, the rates generally grow with temperature. Also, their dependence on the angular coordinate $\theta$ is similar too: (i) on average, the rates tend to increase with $\theta$ and (ii) the functional dependence around $\theta=0$ changes in the same way when one goes from the suprathreshold ($M=3m$) to the subthreshold ($M=m/3$) scalar boson mass. By comparing the results in Figs.~\ref{fig:M3_B4_OmegaS} and \ref{fig:M3_B25_OmegaS}, we also find that the rates are considerably higher in the case of 
stronger field. 

\begin{figure}[t]
 \includegraphics[width=0.31\textwidth]{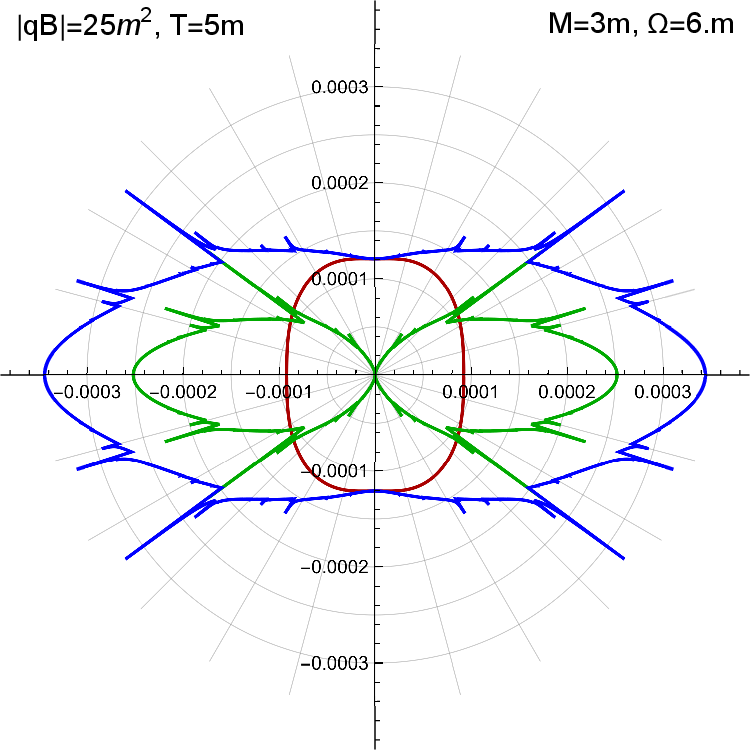}
 \hspace{0.02\textwidth}
 \includegraphics[width=0.31\textwidth]{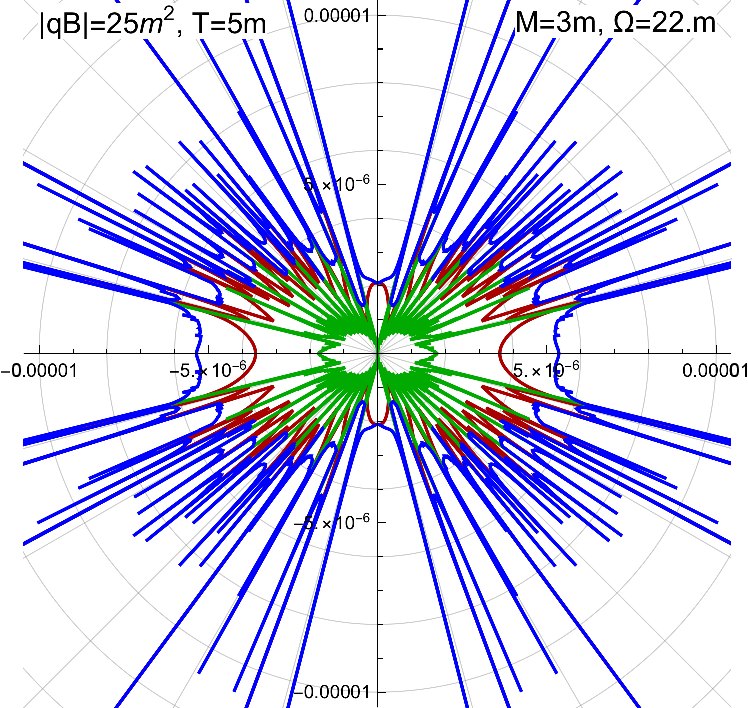}
 \hspace{0.02\textwidth}
  \includegraphics[width=0.31\textwidth]{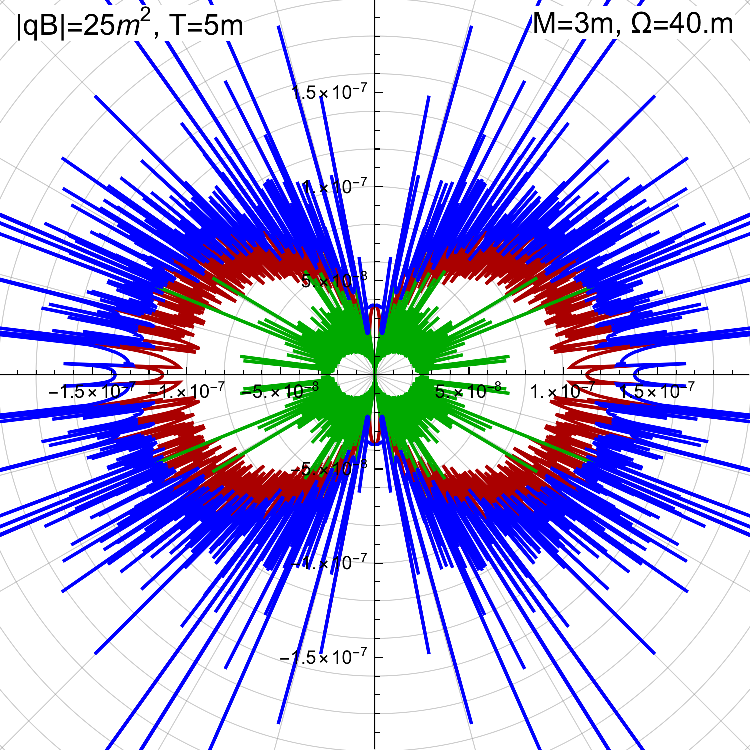}\\[2mm]
 \includegraphics[width=0.31\textwidth]{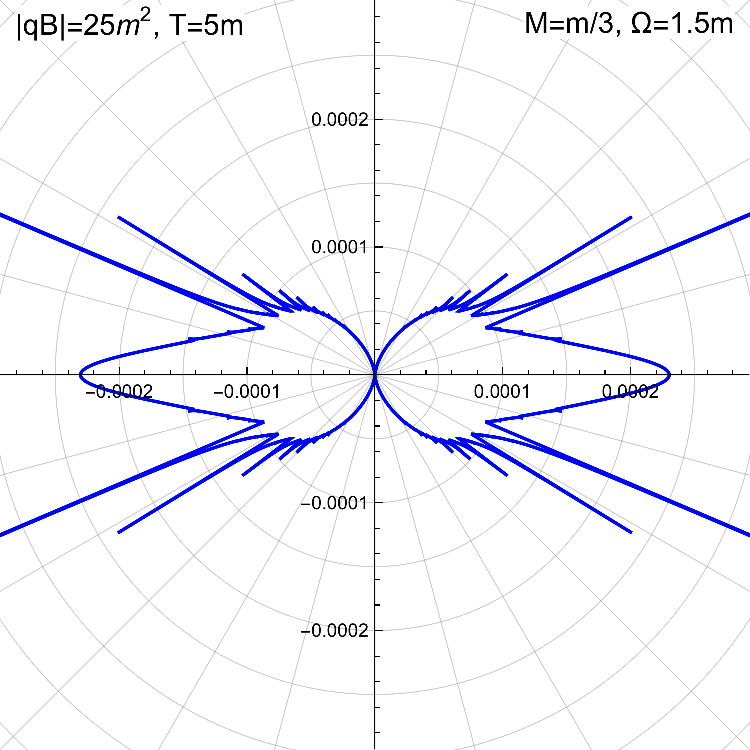}
 \hspace{0.02\textwidth}
 \includegraphics[width=0.31\textwidth]{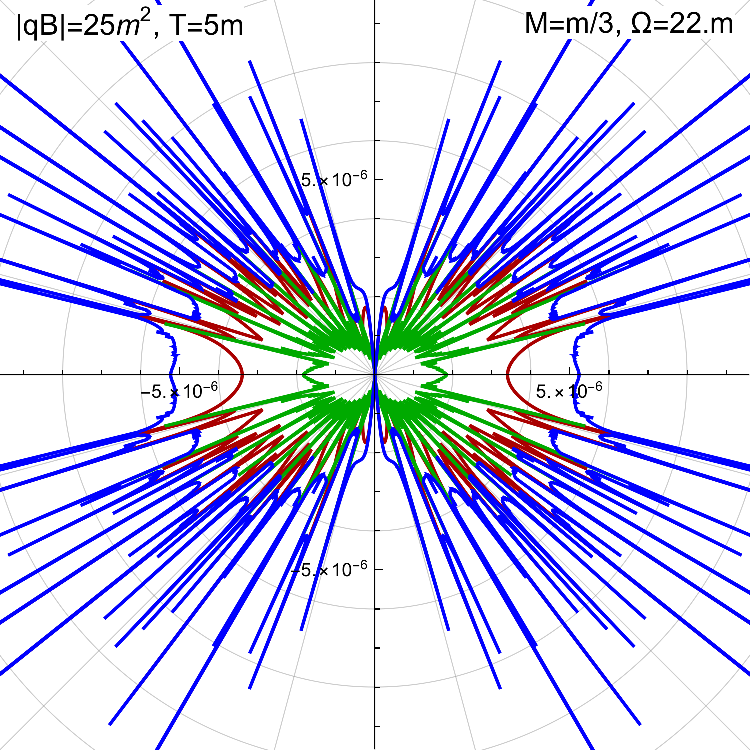}
 \hspace{0.02\textwidth}
  \includegraphics[width=0.31\textwidth]{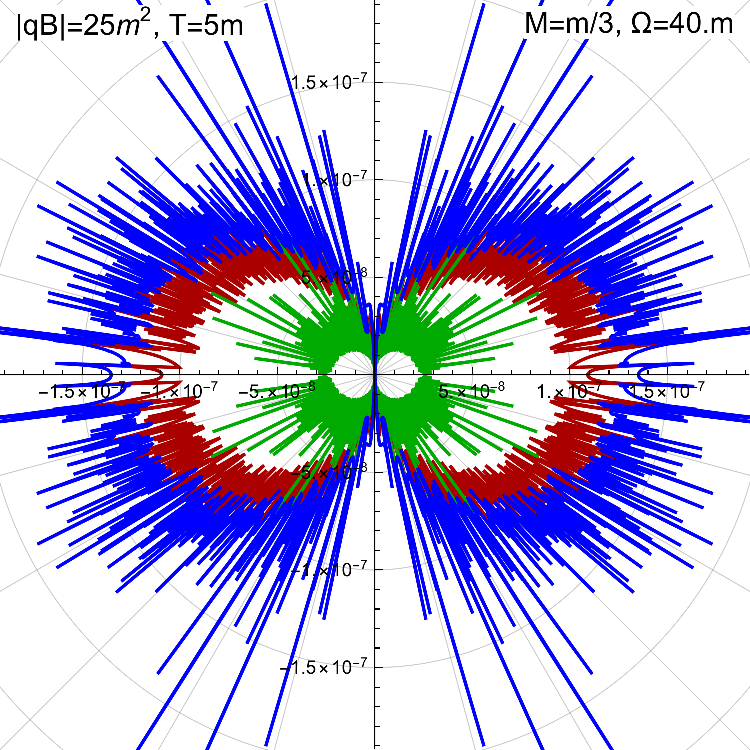}
 \caption{The angular profile of the scalar boson production rates for several different energies and fixed temperature $T=5m$. The magnetic field is $|qB|=25m^2$ and the scalar boson masses are $M=3m$ (top panels) and $M=m/3$ (bottom panels). Each panel contains separate contributions due to annihilation (red lines) and particle-splitting (green lines) processes, as well as the total rates (blue lines).}
 \label{fig:polar_B25_OmegaS}
\end{figure}

The emission profiles and relative contributions of the annihilation and particle-splitting processes in the case of strong field, $|qB|=25m^2$, remain about the same as in the weaker field, $|qB|=4m^2$. Several representative profiles with characteristic butterfly shapes are displayed in six polar plots in Fig.~\ref{fig:polar_B25_OmegaS}. For the scalar mass $M=3m$, the angular distribution of emission is particularly simple at small energies. One of such cases for $\Omega=6m$ is displayed in the top left panel of Fig.~\ref{fig:polar_B25_OmegaS}. At such low energy, the only allowed annihilation processes are those between the lowest Landau levels. As a results, the corresponding rate visualized by the red line has a very smooth profile. Interestingly, it is one of those special cases when the annihilation has a slightly higher probability of producing scalar boson in the direction parallel to the magnetic field. Nevertheless, the particle-splitting processes overcompensate due to their much higher probability to produce scalar bosons in the directions perpendicular to the magnetic field.

There are no surprises in the case of the subthreshold boson mass, $M=m/3$. When $M<\Omega<2m$, again only the particle-splitting processes contribute. It explains why only the blue-line profile is shown in the bottom left panel of Fig.~\ref{fig:polar_B25_OmegaS}, which corresponds to $\Omega=1.5m$. With increasing the energy, the role of annihilation processes grows and they eventually dominate the total rate even for the subthreshold values of the boson mass. In fact, the emission profiles and relative contributions of different processes become very similar at large energies irrespective of the boson mass.

\begin{figure}[t]
 \includegraphics[width=0.45\textwidth]{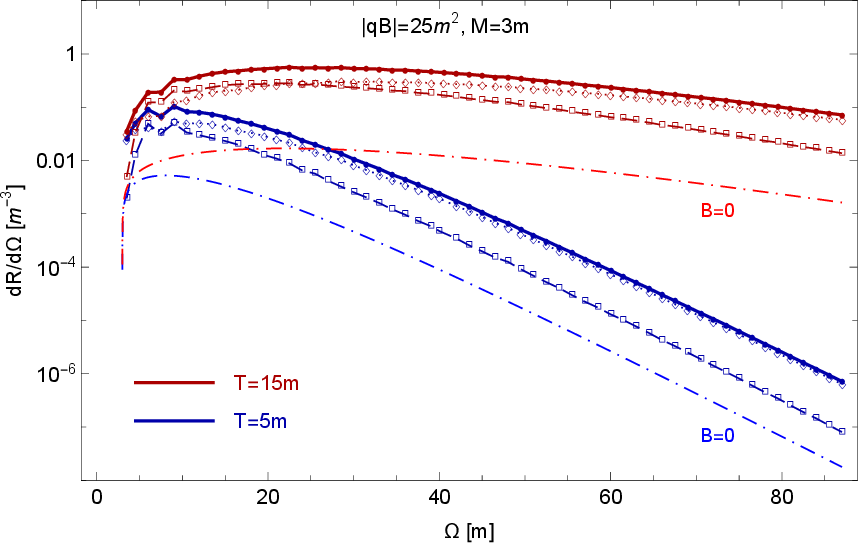}
 \hspace{0.04\textwidth}
 \includegraphics[width=0.45\textwidth]{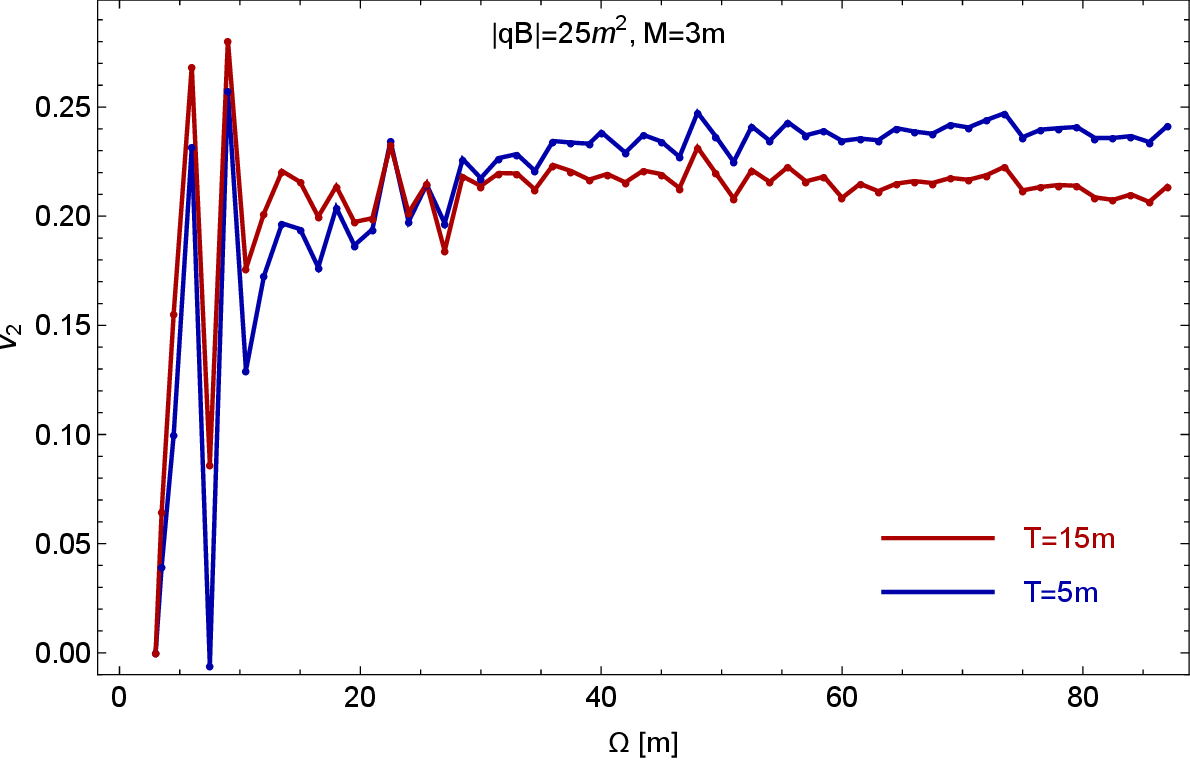}\\[2mm]
 \includegraphics[width=0.45\textwidth]{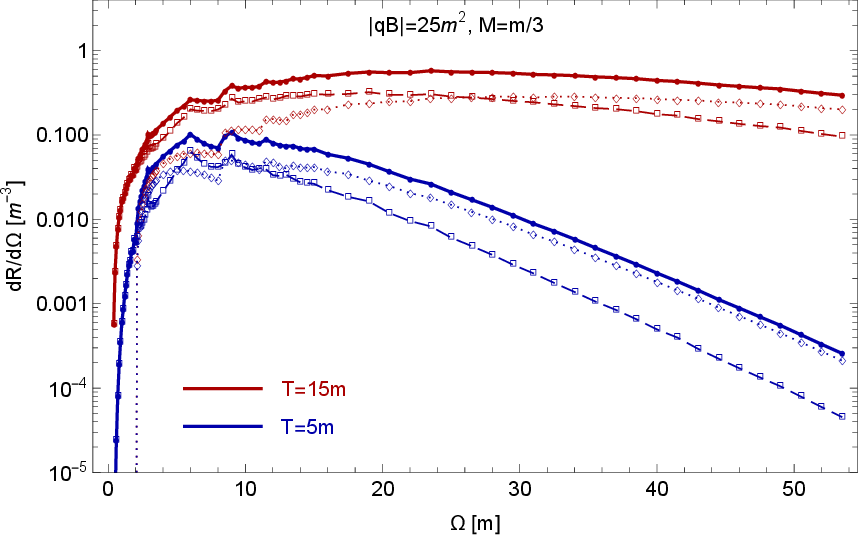}
 \hspace{0.04\textwidth}
 \includegraphics[width=0.45\textwidth]{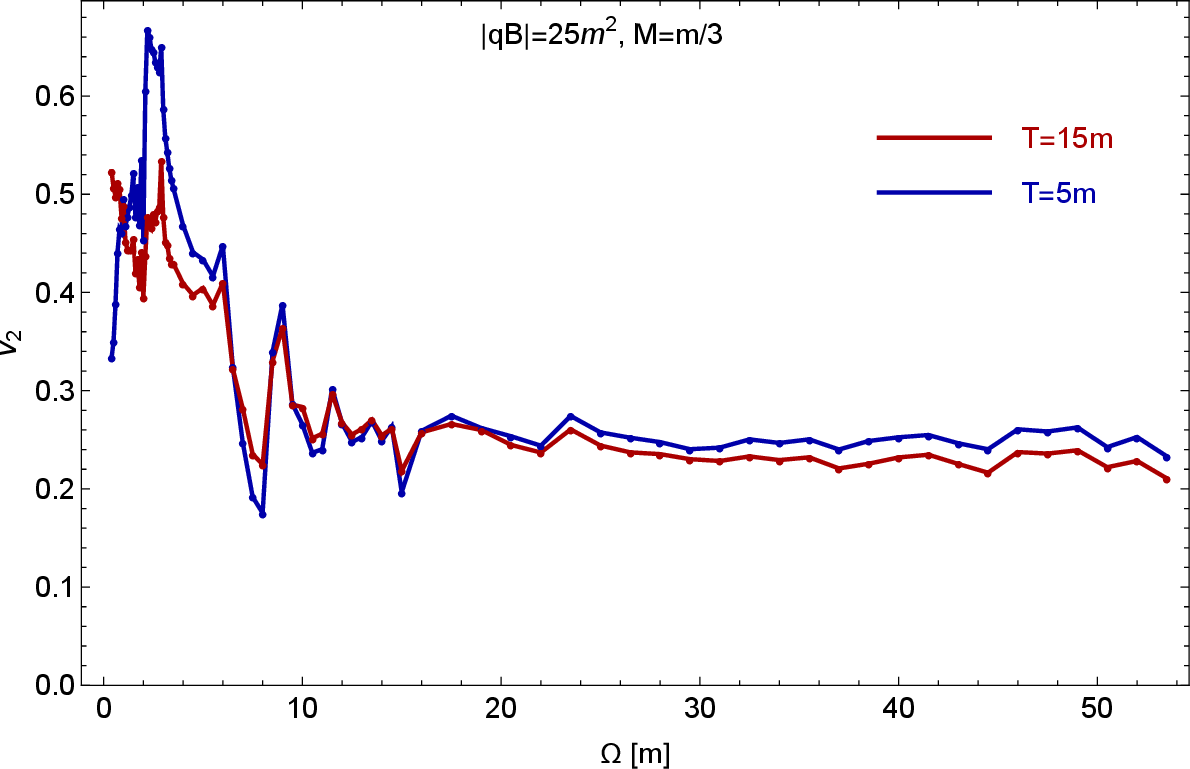}
 \caption{The total rates and ellipticity of scalar boson emission from a magnetized plasma at two different temperatures: $T=5m$ (blue lines) and $T=15m$ (red lines). The magnetic field is $|qB|=25m^2$ and the scalar boson masses are $M=3m$ (top panels) and $M=m/3$ (bottom panels).}
 \label{fig:total-ratesB25}
\end{figure}

For the case of $|qB|=25m^2$, the total rates $dR/d\Omega$ integrated over the angular directions are shown in the two left panels of Fig.~\ref{fig:total-ratesB25}. The two right panels contain the data for the ellipticity measure $v_2$ of the scalar boson production. As before, the results for the lower temperature, $T=5m$, are represented by the blue lines and those for the higher temperature, $T=15m$, are represented by the red lines. Additionally, the filled circles are used as plot markers for the total rate, the open diamonds for annihilation contributions, and the open squares for particle-splitting processes. In the suprathreshold case, $M=3m$, we show additionally the zero-field rates, represented by the dash-dotted lines. (Recall that no nontrivial zero-field limit exists in the subthreshold case with $M=m/3$.)

The energy dependence of the total rates in Fig.~\ref{fig:total-ratesB25} is very similar to the weaker field case in Fig.~\ref{fig:total-ratesB4}. The rates grow with increasing the temperature. The dependence on the scalar boson energy vaguely resembles the black body radiation: the rates grow from zero to its maximum value around the energy $\Omega \sim 1.7 T$ and then decrease by gradually approaching the exponential asymptote, $dR/d\Omega \sim \exp\left(-\Omega/T\right)$. 

The relative contributions of the annihilation and particle-splitting processes can be read off from Fig.~\ref{fig:total-ratesB25} too. While particle-splittings dominate in a range of small energies, $\Omega \lesssim 1.7 T$, the annihilation overwhelms the total rate at high energies, $\Omega \gtrsim 1.7 T$. In the case of larger (smaller) scalar mass $M = 3m$ ($M = m/3$), the corresponding switch of the two regimes occurs at slightly lower (higher) energies. Such a correlation is not surprising since the relative role of annihilation processes is larger (smaller) in the suprathreshold  (subthreshold) case. 

The ellipticity measure $v_2$ of the scalar boson production is again positive and relatively large. Its values are in the same ballpark of $0.2$ to $0.3$. As in the case of the weaker field,  $v_2$ gets slightly suppressed with increasing the temperature. The prominent differences between the cases of $M = 3m$ (suprathreshold) and $M = m/3$ (subthreshold) appear only at small energies, i.e., $\Omega \lesssim 1.7 T$.

\section{Conclusions}
\label{sec.concl}

In this paper, we have derived an analytic expression for the imaginary (absorptive) part of the scalar boson's self-energy within a strongly magnetized relativistic plasma. The model we consider involves a neutral scalar field that interacts with charged fermions through a Yukawa-type coupling. We use the expression for the imaginary part of self-energy to calculate the differential production rate of scalar bosons. In view of the  principle of detailed balance, this same quantity also determines the absorption rate of scalar boson in the magnetized plasma. 

As evident from the explicit expression we have derived, the production rate is determined by three distinct types of processes: particle-splitting ($\psi\rightarrow \psi+\phi $), antiparticle-splitting ($\bar{\psi} \rightarrow \bar{\psi}+\phi $), and particle-antiparticle annihilation ($\psi + \bar{\psi}\rightarrow \phi $). All such processes have been thoroughly analyzed, with careful consideration given to the effects of Landau-level quantization of charged fermions. In the context of a high-temperature relativistic plasma (i.e., $T\gtrsim \sqrt{|qB|}$), our findings reveal that a large number of Landau levels contributes to the rate. In essence, this implies that one cannot rely on the commonly employed lowest Landau level approximation even when the magnetic field is very strong compared to the scale set by the fermion mass. 

The energy dependence of the rates exhibits a resemblance to a black body spectrum, featuring a peak at an intermediate energy level comparable to the plasma's temperature. In our study of several representative cases, we have found that the peak typically occurs at approximately $\Omega \simeq 1.7T$. Also, the rates grow with increasing temperature. The influence of thermal effects can be readily understood. As the temperature rises, the number of occupied positive-energy states and unoccupied negative-energy states grows. It leads to a larger phase space for all the processes contributing to the scalar boson production. 

The rates also exhibit growth with an increasing magnetic field, but the underlying physics is more subtle. One key aspect is the substantial relaxation of momentum conservation constraints provided by the background field. The case in point is the production of bosons through (anti)particle-splitting processes, which are prohibited in the absence of a magnetic field. Additionally, the high degeneracy of Landau levels likely plays a role in enhancing scalar boson production. As in the case of magnetic catalysis \cite{Shovkovy:2012zn}, one may argue that such degeneracy increases the average density of quantum states near small energies. In the case of a hot plasma, this effect translates into an increased phase space for annihilation processes. By comparing the results for two representative field strengths, $|qB|=4m^2$ and $|qB|=25m^2$, as well as for $B=0$, we see that the presence of a magnetic field enhances the average rates.

We also studied in detail the dependence of the differential production rate on the angular coordinate and the scalar boson energy. The butterflylike emission profiles indicate a higher likelihood of boson production in directions perpendicular to the magnetic field. This preference for perpendicular emission is reflected in the ellipticity measure, denoted as $v_2$, which typically assumes positive values in the range of $0.2$ to $0.3$ at high scalar boson energies. At small energies, on the other hand, the values of $v_2$ exhibit greater variability due to energy quantization of the low-lying Landau-level states.  In  this regime, isolated energy thresholds can lead to abrupt changes in the $v_2$ values, rendering this characteristics less informative and of limited utility.

As stated in the Introduction, we do not try to address phenomenological applications in this study. Nevertheless, we cannot help but note that our  findings regarding the production (or decay) rate of scalar bosons may have important implications for cosmology. In particular, they suggest that the primordial magnetic field might exert an even stronger influence on the magnetic warm inflation scenario than previously reported in Refs.~\cite{Piccinelli:2014dya,Piccinelli:2021vbq}. Indeed, now we can fully substantiate the claim that the presence of the magnetic field significantly amplifies the total boson decay rate. Furthermore, the rate far exceeds the contribution from the lowest Landau level, which was employed as an estimate in Ref.~\cite{Piccinelli:2021vbq}.

\begin{acknowledgments}
The visit of J.~J.-U. to Arizona State University was supported by the Universidad Nacional Aut\' onoma de M\' exico through Facultad de Ciencias, CGEP-AANILD, and DGAPA-UNAM under Grant No.~PAPIIT-IN108123. The work of I.~A.~S. was supported by the U.S. National Science Foundation under Grant No.~PHY-2209470. 
\end{acknowledgments}

\appendix

\section{Zero magnetic field}
\label{app.IntegrationB=0}

In this appendix, for comparison purposes, we derive the imaginary part of the scalar boson self-energy in the limit of vanishing magnetic field. Similar results at nonzero temperature can be found in the literature, e.g., see Refs.~\cite{Bastero-Gil:2010dgy,Ho:2015jva}. 

At the leading order, the scalar boson self-energy is given by
\begin{equation}
 \Sigma(k)=ig^2\int\frac{d^4p}{(2\pi)^4}\mbox{Tr}\left[S(p)S(p-k)\right],
 \label{eq.qB=0.self-energy1}
\end{equation}
which is the momentum space representation of a definition analogous to Eq.~(\ref{eq.self-enegry}). In the absence of a background field, the fermion propagator reads
\begin{equation}
 S(p)=i\frac{\slsh{p}+m}{p^2-m^2+i\epsilon}.
 \label{eq.qB=0.fermionpropagator}
\end{equation}
After calculating the Dirac trace and replacing the energy integration with the Matsubara sum, we derive
\begin{equation}
 \Sigma\left(i\Omega_m,\mathbf{k}\right)=4g^2 T\sum_{k=-\infty}^{\infty}\int\frac{d^3p}{(2\pi)^3}\frac{i\omega_n\left(i\omega_n-i\Omega_m\right)-\mathbf{p}\cdot(\mathbf{p}-\mathbf{k})+m^2}{\left[\left(i\omega_n\right)^2-E_p^2\right]\left[\left(i\omega_n-i\Omega_m\right)^2-E_{p-k}^2\right]},
 \label{eq.qB=0.self-energy2}
\end{equation}
where we have introduced the notation for the fermion energies $E_p=\sqrt{\mathbf{p}^2+m^2}$ and $E_{p-k}=\sqrt{(\mathbf{p}-\mathbf{k})^2+m^2}$. 

The zero-field result above is analogous to Eq.~(\ref{eq.MF-VEdiagA9}) in the main text. Similarly, we use Eq.~(\ref{eq.temp.matsubarasum}) to compute the Matsubara sum and arrive at the following result:
\begin{equation}
 \Sigma^R\left(\Omega,\mathbf{k}\right)=g^2 \sum_{\eta,\lambda=\pm1}\int\frac{d^3p}{(2\pi)^3}\frac{n_F\left(E_p\right)-n_F\left(\lambda E_{p-k}\right)}{\lambda E_p E_{p-k}\left(E_p-\lambda E_{p-k}+\eta\Omega+i\eta\epsilon\right)}\left[\lambda E_p E_{p-k}-\mathbf{p}\cdot(\mathbf{p}-\mathbf{k})+m^2\right],
 \label{eq.qB=0.self-energy3}
\end{equation}
where we performed the analytical continuation to Minkowski space by replacing $i\Omega_m\longrightarrow\Omega+i\epsilon$. To separate the real and imaginary parts, we utilize the Sokhotski formula,
\begin{equation}
\frac{1}{E_p-\lambda E_{p-k}+\eta\Omega+i\eta\epsilon} = {\cal P}\frac{1}{E_p-\lambda E_{p-k}+\eta\Omega+i\eta\epsilon} -i\eta\pi \delta\left(E_p-\lambda E_{p-k}+\eta\Omega\right).
\label{eq:Sokhotski}
\end{equation}
Then, the imaginary part of the self-energy is given by
\begin{equation}
 \mbox{Im}\left[\Sigma^R\left(\Omega,\mathbf{k}\right)\right]=-g^2\pi \sum_{\eta,\lambda=\pm1}\int\frac{d^3p}{(2\pi)^3}\frac{n_F\left(E_p\right)-n_F\left(\lambda E_{p-k}\right)}{\eta\lambda E_p E_{p-k}}\left[\lambda E_p E_{p-k}-\mathbf{p}\cdot(\mathbf{p}-\mathbf{k})+m^2\right]\delta\left(E_p-\lambda E_{p-k}+\eta\Omega\right).
 \label{eq.qB=0.self-energy4}
\end{equation}
The remaining integration over the loop momenta can be performed by switching to spherical coordinates,
\begin{eqnarray}
\mbox{Im}\left[\Sigma^R\left(\Omega,\mathbf{k}\right)\right]&=&-g^2\pi \sum_{\eta,\lambda=\pm1}\int_0^\infty\int_{-1}^1\int_0^{2\pi}\frac{\mathbf{p}^2 \ dp  \ dx \ d\varphi}{(2\pi)^3}\frac{n_F\left(E_p\right)-n_F\left(\lambda E_{p-k}\right)}{\eta\lambda E_p E_{p-k}}\nonumber\\
  &&\times\left[\lambda E_p E_{p-k}-\mathbf{p}^2+|\mathbf{p}||\mathbf{k}|x+m^2\right]
  \delta\left(E_p-\lambda E_{p-k}+\eta\Omega\right)\nonumber\\
  &=&-g^2\pi \sum_{\eta,\lambda=\pm1}\int_0^\infty\int_{-1}^1\frac{\mathbf{p}^2 \ dp  \ dx}{(2\pi)^2}\frac{n_F\left(E_p\right)-n_F\left(\lambda E_{p-k}\right)}{\eta\lambda E_p E_{p-k}}\nonumber\\
  &&\times\left[\lambda E_p E_{p-k}-\mathbf{p}^2+|\mathbf{p}||\mathbf{k}|x+m^2\right]|E_p+\eta\Omega|\frac{\delta\left(x-x_0\right)}{|\mathbf{p}||\mathbf{k}|},
 \label{eq.qB=0.self-energy5}
\end{eqnarray}
where we used the properties of the Dirac $\delta$-function and took into account the following solution to the energy-conservation equation:
\begin{equation}
 x_0=-\frac{\Omega^2-|\mathbf{k}|^2+2\eta E_p\Omega}{2|\mathbf{p}||\mathbf{k}|}.
\end{equation}
Changing the integration variable from $p$ to energy $E_p$, we derive
\begin{equation}
  \mbox{Im}\left[\Sigma^R\left(\Omega,\mathbf{k}\right)\right]=-\frac{g^2\pi}{(2\pi)^2} \int_{E_{-}}^{E_{+}}dE_p\frac{n_F\left(E_p\right)-n_F\left(E_p-\Omega\right)}{|\mathbf{k}|}\left(2m^2-\frac{\Omega^2-|\mathbf{k}|^2}{2}\right)\Theta\left(\Omega-E_p\right)\Theta\left(\Omega^2-|\mathbf{k}|^2-4m^2\right),
 \label{eq.qB=0.self-energy6}
\end{equation}
where the integration limits are defined by
\begin{equation}
 E_{\pm}\equiv\frac{\Omega}{2}\pm\frac{|\mathbf{k}|}{2}\sqrt{1-\frac{4m^2}{\Omega^2-|\mathbf{k}|^2}}.
 \label{eq.qB=0.Energies}
\end{equation}
These were obtained by requiring that $-1<x_0<1$. After integrating over the energy, the final result reads 
\begin{equation}
  \mbox{Im}\left[\Sigma^R\left(\Omega,\mathbf{k}\right)\right]=-\frac{g^2}{8\pi}\left(\Omega^2-|\mathbf{k}|^2-4m^2\right)\left[\sqrt{1-\frac{4m^2}{\Omega^2-|\mathbf{k}|^2}}+\frac{2T}{|\mathbf{k}|}\ln\left(\frac{1+e^{-\beta E_{+}}}{1+e^{-\beta E_{-}}}\right)\right]\Theta\left(\Omega^2-|\mathbf{k}|^2-4m^2\right).
 \label{eq.qB=0.self-energy7}
\end{equation}

\bibliographystyle{apsrev4-1}
\bibliography{BIBLIO2}

\end{document}